\documentclass[12pt,a4paper]{article}

\usepackage{amsmath,amssymb,amsthm,mathtools}
\usepackage{lmodern}
\usepackage[T1]{fontenc}

\usepackage{geometry,booktabs,array,xcolor,enumitem}
\usepackage{bm}
\usepackage{hyperref}
\newtheorem{theorem}{Theorem}[section]

\theoremstyle{definition}

\newtheorem{remark}[theorem]{Remark}

\newcommand{\eps}{\epsilon}

\newcommand{\om}{\omega}
\newcommand{\ka}{\kappa}
\newcommand{\al}{\alpha}
\newcommand{\De}{\Delta}
\newcommand{\de}{\delta}

\newcommand{\Oc}{\mathcal{O}}

\newcommand{\svn}{\sqrt{7}}

\newcommand{\zFG}{z_{\mathrm{FG}}}
\newcommand{\rh}{r_{h}}

\newcommand{\CFT}{\mathrm{CFT}}

\newcommand{\vev}[1]{\langle #1 \rangle}

\newcommand{\hatP}{\hat{P}_{\nu}}

\newcommand{\mR}{m_{\mathrm{R}}}

\begin{document}

\title{\large\textbf{
  Perturbative Hairy Black Branes in the $G_{2}$-Invariant Sector of Dyonic ISO(7) Gauged Supergravity}}

\author{Sangheon Yun\\
  \small IndigoWave, Center for Quantum Spacetime, Sogang University,\\
  \small 35 Baekbeom-ro, Mapo-gu, Seoul 04107, Republic of Korea}
\date{}
\maketitle
\vspace*{8mm}

\begin{abstract}
We construct perturbative black-brane solutions carrying neutral scalar hair in the $G_{2}$-invariant truncation of four-dimensional dyonic $\mathcal{N}=8$ ISO(7) gauged supergravity, an exact consistent truncation of massive type~IIA supergravity on $S^{6}$.
Expanding around the $G_{2}$-symmetric AdS$_{4}$-Schwarzschild black brane in powers of a dimensionless scalar-charge parameter $\eps$, we show that the perturbation hierarchy reduces, order by order, to a nested sequence of inhomogeneous Legendre equations (demonstrated through second order).
The first-order scalar satisfies a P\"{o}schl--Teller equation with exact parameter $\nu(\nu+1)=2/3$, giving $\nu=-\frac{1}{2}+\frac{\sqrt{33}}{6}$, and the first-order metric is fixed algebraically by the same function, $A^{(1)}=-\tfrac{\sqrt{7}}{2}\phi^{(1)}$, so the entire $\mathcal{O}(\eps)$ sector is one Legendre function.
An identity $\nu+1=\Delta_{+}/3$, with $\Delta_{+}=(3+\sqrt{33})/2\approx 4.372$ the conformal dimension of the dual scalar operator, relates the P\"{o}schl-Teller parameter to the AdS/CFT dictionary via the nonlinear Fefferman-Graham map $z \propto z_{\mathrm{FG}}^{3}$.
The $\mathcal{O}(\eps^{2})$ back-reaction yields explicit corrections to the renormalized free energy and Bekenstein-Hawking entropy density at fixed temperature, consistent with the first law; equivalently, at fixed entropy or energy density it \emph{lowers} the Hawking temperature by $\delta T/T=\mathcal{O}(\eps^{2})$.
For the leading transport coefficients the shear viscosity saturates $\eta/s=1/(4\pi)$, while the Eling-Oz horizon formula gives a nonzero bulk viscosity $\zeta/\eta=\bigl(\tfrac{\Delta_{+}-3}{2}\bigr)^{2}\eps^{2}=\tfrac{3(7-\sqrt{33})}{8}\eps^{2}\approx0.471\,\eps^{2}$.
With $c_{s}^{2}=\tfrac12-\mathcal O(\eps^{2})<\tfrac12$, this indicates a softening of the equation of state.
\end{abstract}

\vspace{6em}

\vspace*{5mm}
\noindent E-mail address: \texttt{sangheon.yun@gmail.com}

\tableofcontents
\newpage

\section{Introduction}
\label{sec:intro}

\subsection{The AdS/CFT correspondence and holographic transport}
\label{sec:intro_adscft}

The gauge/gravity duality, first proposed by Maldacena~\cite{Maldacena:1997re} and given its precise field-theoretic formulation by Gubser, Klebanov, and Polyakov~\cite{Gubser:1998bc} and by Witten~\cite{Witten:1998qj}, asserts an exact equivalence between a $(D+1)$-dimensional theory of quantum gravity (or string theory) on anti-de~Sitter space AdS$_{D+1}$ and a $D$-dimensional
conformal field theory (CFT$_{D}$), with $d=D-1$ spatial dimensions, living on its conformal boundary.
In its most controlled incarnation, the duality relates type~IIB string theory on AdS$_{5}\times S^{5}$ to $\mathcal{N}=4$ super Yang--Mills theory (SYM) with gauge group $SU(N)$ in the large-$N$, strong-coupling limit~\cite{Maldacena:1997re}.
The dictionary between the two descriptions is encoded in the GKPW prescription~\cite{Gubser:1998bc,Witten:1998qj}: the on-shell bulk action, evaluated with appropriate Dirichlet boundary conditions, equals the generating functional of connected correlation functions in the dual CFT.

One of the most striking early applications of this correspondence was the calculation of real-time transport coefficients in the strongly coupled $\mathcal{N}=4$ SYM plasma.
Policastro, Son, and Starinets~\cite{Policastro:2001yc} showed that the shear viscosity of the plasma is related to the absorption cross section of a minimally coupled graviton on the AdS$_{5}$ black-brane background, obtaining $\eta=\pi N^{2}T^{3}/8$ at strong coupling.
Combined with the entropy density $s=\pi^{2}N^{2}T^{3}/2$ of the AdS$_{5}$-Schwarzschild black brane, this gives the celebrated ratio
\begin{equation}\label{eq:etas}
  \frac{\eta}{s} = \frac{1}{4\pi}\,,
\end{equation}
first derived in~\cite{Policastro:2001yc} and elevated to a rigorous lower bound--the \emph{KSS bound}--by Kovtun, Son, and Starinets~\cite{Kovtun:2004de}.
The universality of~\eqref{eq:etas} within two-derivative Einstein gravity was proved by Buchel and Liu~\cite{Buchel:2003tz}.

The ratio~\eqref{eq:etas} takes its minimal value precisely when conformal invariance is preserved: for a relativistic conformal fluid, $\zeta=0$ exactly~\cite{Romatschke:2009im}, where $\zeta$ denotes the bulk viscosity.
Once conformal symmetry is broken, however, the bulk viscosity becomes nonzero.
For a holographic plasma dual to a bulk theory in $D+1$ dimensions in which an operator $\mathcal{O}$ of conformal dimension $\Delta$ is sourced by a coupling $J$, the bulk viscosity is controlled by the horizon value of the dual scalar through the Eling-Oz horizon formula~\cite{Gubser:2008sz,Buchel:2008mf,Gubser:2008ny,Eling:2011ms}
\begin{equation}\label{eq:zeta_ward}
  \frac{\zeta}{\eta} =
  \left(\frac{d\phi_{H}}{d\ln s}\right)^{\!2}_{\!J},
\end{equation}
which is exact for a single canonically normalized bulk scalar and vanishes precisely at the conformal point (constant $\phi_{H}$).
The holographic computation of $\zeta$ therefore requires knowing the equation of state of the plasma as a function of the source $J$, which in turn requires constructing the hairy black-brane solution that back-reacts to the scalar perturbation.
The main goal of the present paper is to construct such hairy black branes analytically, to second order in the scalar charge, in a fully explicit embedding into string theory.

\subsection{Massive type~IIA supergravity and the ISO(7) gauging}
\label{sec:intro_IIA}

\begin{sloppypar}
Recent years have witnessed remarkable progress in constructing explicit four-dimensional gauged supergravities from string/M-theory compactifications with a clear microscopic interpretation.
Particularly important is the dyonic ISO(7) gauged $\mathcal{N}=8$ supergravity discovered by Dall'Agata, Inverso, and Marrani~\cite{Dall'Agata:2014bb}.
Guarino, Jafferis, and Varela~\cite{Guarino:2015jca} showed that this theory has a string-theory origin as an exact consistent truncation of massive type~IIA (Romans) supergravity~\cite{Romans:1985tz} on the six-sphere $S^{6}$:
\end{sloppypar}
\begin{equation}
  \text{mIIA on }S^{6} \;\longrightarrow\;
  \text{dyonic ISO(7) } \mathcal{N}=8 \text{ SUGRA}_{4}.
\end{equation}
This is one of the very few known consistent truncations of a ten-dimensional string theory to a maximal gauged supergravity in four dimensions.
The gauge group ISO(7)$=\mathrm{SO}(7)\ltimes\mathbb{R}^{7}$ combines the isometry group $\mathrm{SO}(7)$ of $S^{6}$ with an additional $\mathbb{R}^{7}$ gauge symmetry associated with the
Romans mass deformation; the Romans mass $\mR=F_{0}$ itself controls the dyonic deformation that is absent in purely electric gaugings.

The ISO(7) theory has a remarkably rich landscape of AdS$_{4}$ critical points, all of which have been classified by Borghese, Guarino, and Roest~\cite{Borghese:2012qm}.
Several of these vacua admit explicit ten-dimensional uplifts together with dual Chern-Simons-matter descriptions: for instance, the $\mathcal{N}=3$ AdS$_{4}$ point of the SO(3)$\times$SO(3) sector uplifts to a warped $\mathrm{AdS}_{4}\times X_{6}$ solution of massive IIA~\cite{Pang:2015rwd}, its dual $\mathcal{N}=3$ Chern-Simons-matter theory has been identified through
matching of the localized free energy~\cite{Pang:2015vna}, and the Kaluza-Klein spectra of these holographic Chern-Simons theories exhibit notable universality~\cite{Pang:2017omp}.
Among them, the two $G_{2}$-invariant critical points play a central role:
\begin{itemize}[leftmargin=1.8em]
  \item \textbf{The $\mathcal{N}=1$ supersymmetric $G_{2}$ vacuum.}
    This was first found by Warner~\cite{Warner:1983vz} in the $\mathcal{N}=8$ SO(7) gauged supergravity and identified in the ISO(7) theory in~\cite{Guarino:2015qaa}.
    It preserves $\mathcal{N}=1$ supersymmetry and is dual to the IR fixed point of a mass deformation of the ABJM theory~\cite{Aharony:2008ug,Guarino:2015jca}.

  \item \textbf{The non-supersymmetric $G_{2}$ vacuum.}
    Discovered in~\cite{Borghese:2012qm}, this critical point has no supersymmetric completion.
    Its perturbative stability--the fact that all 70 scalar masses satisfy the Breitenlohner-Freedman (BF) bound~\cite{Breitenlohner:1982bm}--was proved analytically by Guarino, Malek, and Samtleben~\cite{Guarino:2020ltm}, making it one of the rare examples of a perturbatively stable non-supersymmetric AdS vacuum in string theory.
\end{itemize}

\subsection{Scalar hair in holographic black branes}
\label{sec:intro_hair}

\emph{Scalar hair} in the context of AdS gravity refers to a black-brane (or black-hole) solution in which the bulk scalar field $\phi$ is nontrivial: it departs from its constant critical-point value and carries a radially dependent profile that sources the dual scalar operator $\mathcal{O}_{\phi}$.
The earliest systematic studies of hairy AdS black holes were carried out by Hertog and Maeda~\cite{Hertog:2004ns,Hertog:2004dr}, who constructed numerical hairy black-hole solutions in $\mathcal{N}=8$ gauged supergravity and mapped out their thermodynamic phase structure.
These solutions break conformal invariance \emph{explicitly}: the scalar acquires a nonzero source $J\neq 0$, and the free energy per unit area satisfies $\mathcal{F}=\mathcal{F}(T,J)$.

A qualitatively different phenomenon is \emph{spontaneous} scalar hair (holographic superconductivity), discovered by Gubser~\cite{Gubser:2008px} and by Hartnoll, Herzog, and Horowitz~\cite{Hartnoll:2008vx,Hartnoll:2008kx}.
There, a $U(1)$-charged scalar condenses below a critical temperature, breaking the $U(1)$ gauge symmetry spontaneously and producing a hairy black hole with nonzero normalizable mode ($\vev{\mathcal{O}}\neq 0$) and vanishing source ($J=0$).

In the present paper we consider \emph{explicit} scalar hair sourced by a neutral scalar field $\phi$ in the $G_{2}$-invariant ISO(7) sector.
The scalar $\phi$ is dual to an \emph{irrelevant} operator $\mathcal{O}_{\phi}$ of dimension $\Delta_{+}=(3+\sqrt{33})/2 \approx 4.372>3$ (see Section~\ref{sec:bg}), and turning on a nonzero source $J\neq 0$ corresponds to adding an \emph{irrelevant} deformation to the SCFT$_{3}$ Lagrangian:
\begin{equation}\label{eq:defLag}
  \mathcal{L}_{\CFT} \;\to\; \mathcal{L}_{\CFT} + J\,\mathcal{O}_{\phi}.
\end{equation}
At finite temperature $T$, this deformation shifts the free energy by $\delta\mathcal{F}\sim J^{2}T^{2\Delta_{+}-3}+\cdots$, which in turn produces a nonzero bulk viscosity via~\eqref{eq:zeta_ward}.

Because $\Delta_{+}>3$, the source mode grows toward the boundary, $\phi\sim J\,r^{\Delta_{+}-3}$, so at finite deformation the solution does not obey standard asymptotically AdS$_{4}$ boundary conditions.
This is a property of any irrelevant coupling rather than a pathology of the model: such a deformation defines the theory as an effective description, organized order by order in $J$ exactly as a
non-renormalizable coupling is organized in effective field theory.
The $\eps$-expansion used here implements that organization--at each order the boundary conditions are well posed and the holographic counterterms close on a finite set (Appendix~\ref{app:holoren})--and its domain of validity is quantified in Sec.~\ref{sec:eff_dim}.

Analytic hairy black-brane solutions in AdS are rare.
Known examples include the polynomial potential~\cite{Anabalon:2012sn,Anabalon:2012sra,Acena:2012mr} and the exponential potential(s)~\cite{Anabalon:2012sra,Feng:2013tza,Yun:2026exact,Cadoni:2011sm,Cadoni:2012uf}.
The common feature of these solutions is that the potential allows a factorization of the radial equation.
What distinguishes the present work is that our solution is embedded in a fully explicit string-theory reduction (mIIA on $S^6$) and that the key algebraic structure--the
reduction of the linearized equation for $\phi^{(1)}$ to an \emph{associated Legendre equation} with rational P\"{o}schl-Teller parameter $\nu(\nu+1)=2/3$--follows
directly and unavoidably from the ISO(7) scalar potential, without any \emph{ad hoc} ansatz on the potential.
A complementary analytic hairy black brane, arising from a neutral-scalar truncation of type~IIA supergravity relevant to thermal ABJM dynamics and describing an \emph{effectively marginally relevant} operator with logarithmic RG flow, was analyzed by the present author in~\cite{Yun:2026exact,Bak:2010yd,Yun:2010}; there the deformation is exactly solvable, the geometry reaches the $AdS_4$ fixed point exactly in the infrared at zero temperature, and the plasma becomes conformal as $T/\Lambda \rightarrow 0$ with corrections of relative order $(T/\Lambda)^3$, whereas the present deformation is treated perturbatively, with the P\"{o}schl-Teller/Legendre structure emerging exactly order by order.

\subsection{The P\"{o}schl-Teller potential and its role here}
\label{sec:intro_PT}

The P\"{o}schl-Teller (PT) equation~\cite{Poeschl:1933zz}
\begin{equation}\label{eq:PT_intro}
  -\frac{d^{2}\psi}{dz^{2}}
  +\frac{\om^{2}\,\nu(\nu+1)}{\sinh^{2}(\om z)}\,\psi = 0
\end{equation}
appears throughout quantum mechanics and mathematical physics.
In the AdS/CFT context the same potential governs the quasi-normal mode (QNM) spectrum of scalars in AdS$_{3}$~\cite{Birmingham:2001pj} and appears as the radial equation for minimal scalars in the BTZ black hole~\cite{Banados:1992wn}.
The two linearly independent solutions of the PT equation are the associated Legendre functions $P_{\nu}(\coth\om z)$ and $Q_{\nu}(\coth\om z)$, which have well-studied asymptotic behaviors.

In our context, the PT equation~\eqref{eq:PT_intro} emerges directly from linearizing the Klein-Gordon equation for $\phi$ around the $G_{2}$ AdS$_{4}$-Schwarzschild background.
The exact rationality of $\nu(\nu+1)=2/3$ is an intrinsic property of the ISO(7) model--specifically, it follows from $m_{\phi}^{2}l^{2}/9=2/3$, where $m_{\phi}^{2}=6/l^{2}$ is the scalar mass at the $G_{2}$ critical point.
We exploit this exact solvability to build the perturbation theory to second order.

\subsection{Structure of the paper and summary of main results}
\label{sec:intro_summary}

The paper is organized as follows.
Section~\ref{sec:model} introduces the $G_{2}$-invariant ISO(7) supergravity model, including the action, scalar potential, and reduction to a two-function system of ODEs.
Section~\ref{sec:bg} analyzes the $G_{2}$ critical point, constructs the AdS$_{4}$-Schwarzschild black brane in the physical radial coordinate~$r$, and computes its zeroth-order thermodynamics.
Section~\ref{sec:pert} develops the perturbative expansion in the scalar-charge parameter $\eps$ to second order.
Section~\ref{sec:holo} extracts the holographic dictionary, derives the Fefferman-Graham expansion, and collects all transport coefficients.
Section~\ref{sec:numerics} evaluates the solutions explicitly--the regular kernels, the perturbation profiles and the holographic observables--and derives the thermodynamic response of the
deformed plasma.
Section~\ref{sec:disc} discusses physical implications, stability considerations, and future directions.
Appendix~\ref{app:legendre} collects useful properties of the Legendre functions, and Appendix~\ref{app:holoren} carries out the holographic renormalization of the on-shell action.

Our principal results are:

\paragraph{Result~1: Exact P\"{o}schl-Teller reduction.}
The first-order scalar perturbation satisfies (in $r$):
\begin{equation}\label{eq:res1}
  \phi^{(1)\prime\prime}
  + \frac{4r^{3}-\rh^{3}}{r(r^{3}-\rh^{3})}\,\phi^{(1)\prime}
  - \frac{6r}{r^{3}-\rh^{3}}\,\phi^{(1)} = 0,
\end{equation}
with exact solution $\phi^{(1)}(r)=P_{\nu}(2r^{3}/\rh^{3}-1)$, $\nu(\nu+1)=2/3$.
The first-order metric perturbation is then fixed algebraically by the same function,
\begin{equation}\label{eq:res1b}
  A^{(1)}(r) = -\frac{\svn}{2}\,\phi^{(1)}(r),
\end{equation}
so the entire $\mathcal{O}(\eps)$ sector is a single Legendre function.

\paragraph{Result~2: Exact holographic identity.}
\begin{equation}\label{eq:res2}
  \nu + 1 = \frac{\Delta_{+}}{3},
  \qquad
  \Delta_{+} = \frac{3+\sqrt{33}}{2},
\end{equation}
relating the P\"{o}schl-Teller parameter to the conformal dimension of the dual operator via the nonlinear FG map $z\sim z_{\mathrm{FG}}^{3}$.

\paragraph{Result~3: Transport coefficients.}
The shear viscosity saturates $\eta/s=1/(4\pi)$ exactly, while the bulk viscosity is $\zeta/\eta = \tfrac{3(7-\sqrt{33})}{8}\,\eps^{2} \approx 0.471\,\eps^{2}$.

\paragraph{Result~4: Equation of state and thermodynamic response.}
At $\mathcal{O}(\eps^{2})$ the dual plasma becomes nonconformal: $\varepsilon-2p=(\De_{+}-3)\,J\langle\mathcal{O}_{\phi}\rangle/l^2 \propto T^{\sqrt{33}}$ at fixed source, the speed of sound softens to $c_{s}^{2}<\tfrac12$, and at fixed entropy or energy density the Hawking temperature is \emph{lowered}, $\delta T/T=\mathcal{O}(\eps^{2})<0$.
In dimensionless form--$(\varepsilon-2p)/p$, $c_{s}^{2}-\tfrac12$ and $\delta T/T$--all three are governed by the single combination $\chi_{0}J^{2}\rh^{\sqrt{33}-3}\propto T^{\sqrt{33}-3}$, and therefore switch off as $T\to0$, where conformality is restored.

We keep Newton's constant $G_{4}$ explicit throughout and use metric signature $(-,+,+,+)$.

\section{The \texorpdfstring{$G_{2}$}{G2}-Invariant ISO(7) Supergravity Model}
\label{sec:model}

\subsection{ISO(7) gauged $\mathcal{N}=8$ supergravity}
\label{sec:model_iso7}

Dyonic ISO(7) gauged $\mathcal{N}=8$ supergravity in four dimensions~\cite{Dall'Agata:2014bb,Guarino:2015jca} contains the $\mathcal{N}=8$ graviton multiplet, which comprises the
metric $g_{\mu\nu}$, 8 gravitini $\psi_{\mu}^{I}$, 28 vectors $A_{\mu}^{IJ}$, 56 gaugini $\chi^{IJK}$, and 70 real scalars parameterizing the coset space $E_{7(7)}/SU(8)$.
The gauge group is the semi-direct product ISO(7)$=\mathrm{SO}(7)\ltimes\mathbb{R}^{7}$, with electric gauge coupling $g_{c}$ for the $\mathrm{SO}(7)$ factor and Romans mass $\mR$ for the $\mathbb{R}^{7}$ factor.

Under the $G_{2}\subset\mathrm{SO}(7)$ subgroup, the 70 scalars decompose into $G_{2}$-singlets and non-singlets~\cite{Borghese:2012qm}.
There are exactly two real $G_{2}$-singlet scalars: the dilaton $\phi$ and the axion $\chi$.
These two scalars form the \emph{$G_{2}$-invariant sector}, and their dynamics are controlled by a two-scalar potential $V(\phi,\chi)$~\cite{Guarino:2015qaa,Borghese:2012qm}.
All remaining 68 scalars are nontrivial representations of $G_{2}$ and may be consistently set to zero.

\subsection{Consistent truncation to \texorpdfstring{$\chi=0$}{chi=0}}
\label{sec:model_trunc}

In this paper we further truncate to $\chi=0$.
The scalar $\chi$ transforms as a pseudoscalar under the $\mathbb{Z}_{2}$ parity symmetry $\chi\to-\chi$ of the full $G_{2}$-invariant potential:
\begin{equation}\label{eq:Z2}
  V(\phi,-\chi) = V(\phi,\chi)\quad\forall\;\phi,
\end{equation}
so $\partial V/\partial\chi|_{\chi=0}=0$ for all $\phi$.
This makes $\chi=0$ an \emph{exact consistent truncation}: any solution with $\chi=0$ initial data remains at $\chi=0$.

With $\chi=0$, the bosonic action in our conventions is
\begin{equation}\label{eq:action}
  S = \frac{1}{16\pi G_{4}}\int d^{4}x\,\sqrt{-g}\!\left[ R - \frac{1}{2}(\nabla\phi)^{2} - V(\phi) \right],
\end{equation}
with the two-exponential scalar potential
\begin{equation}\label{eq:V}
  V(\phi) = -\frac{35g_{c}^{2}}{2}\,e^{-\phi/\svn} + \frac{\mR^{2}}{2}\,e^{-\svn\phi}.
\end{equation}
The two exponential terms originate from the $\mathrm{SO}(7)$-gauge and Romans-mass contributions to the $\mathcal{N}=8$ scalar potential, respectively.

The equations of motion derived from~\eqref{eq:action} are the Einstein equations
\begin{equation}\label{eq:Einstein}
  G_{\mu\nu} = \frac{1}{2}\nabla_{\mu}\phi\nabla_{\nu}\phi - \frac{1}{4}g_{\mu\nu}(\nabla\phi)^{2} - \frac{1}{2}g_{\mu\nu}V(\phi)
\end{equation}
and the Klein-Gordon equation
\begin{equation}\label{eq:KG}
  \Box\phi = \frac{\partial V}{\partial\phi} = \frac{5\svn g_{c}^{2}}{2}\,e^{-\phi/\svn} - \frac{\mR^{2}\svn}{2}\,e^{-\svn\phi}.
\end{equation}

\subsection{Planar metric ansatz and reduction to ODEs}
\label{sec:model_ansatz}

We seek static, planar-symmetric solutions (black branes) with metric
\begin{equation}\label{eq:metric_ansatz}
  ds^{2} = -e^{2C(z)}\,dt^{2} + e^{2E(z)}\,dz^{2} + e^{2H(z)}\!\left(dx^{2}+dy^{2}\right),
\end{equation}
where the radial gauge $E=C+2H$ is imposed, $z=0$ is the conformal boundary, and $z=z_{h}$ is the horizon.

It is convenient to introduce the combinations
\begin{equation}\label{eq:ABdef}
  A \equiv C + 2H - \frac{\svn}{2}\,\phi,\qquad B \equiv C + 2H - \frac{\phi}{2\svn},
\end{equation}
motivated by the structure of the potential~\eqref{eq:V}.
Note that $B = A + \frac{3}{\svn}\,\phi$.
Substituting the ansatz into the Einstein equations~\eqref{eq:Einstein} and the Klein-Gordon equation~\eqref{eq:KG}, one finds (with primes denoting $d/dz$):
\begin{align}
  A'' &= \frac{35g_{c}^{2}}{2}\,e^{2B} + \mR^{2}\,e^{2A}, \label{eq:Aeom}\\
  B'' &= 25g_{c}^{2}\,e^{2B} - \frac{\mR^{2}}{2}\,e^{2A}, \label{eq:Beom}\\
  \phi'' &= \frac{5\svn g_{c}^{2}}{2}\,e^{2B} - \frac{\mR^{2}\svn}{2}\,e^{2A}. \label{eq:phieom}
\end{align}
Additionally, one has the constraint $C=H-\om z+\text{const}$, where $\om$ is a constant set by the asymptotic conditions.

\section{The \texorpdfstring{$G_{2}$}{G2}-Symmetric AdS\texorpdfstring{$_{4}$}{4}-Schwarzschild Black Brane}
\label{sec:bg}

\subsection{The \texorpdfstring{$G_{2}$}{G2} critical point}
\label{sec:bg_cp}

A constant scalar $\phi=\phi_{0}$ solves~\eqref{eq:phieom} if and only if $V'(\phi_{0})=0$:
\begin{equation}\label{eq:crit}
  \frac{5\svn g_{c}^{2}}{2}\,e^{-\phi_{0}/\svn} = \frac{\mR^{2}\svn}{2}\,e^{-\svn\phi_{0}}\quad\Longleftrightarrow\quad e^{6\phi_{0}/\svn} = \frac{\mR^{2}}{5g_{c}^{2}} \equiv \al.
\end{equation}
The AdS$_{4}$ radius $l$ is determined by $V(\phi_{0})=-6/l^{2}$:
\begin{equation}\label{eq:phi0_vals}
  e^{-\phi_{0}/\svn} = \frac{2}{5g_{c}^{2}l^{2}}, \qquad e^{-\svn\phi_{0}} = \frac{2}{\mR^{2}l^{2}}.
\end{equation}
The scalar mass squared at the critical point is
\begin{equation}\label{eq:mphi}
  m_{\phi}^{2} \equiv V''(\phi_{0}) = \frac{6}{l^{2}},
\end{equation}
which lies above the Breitenlohner-Freedman bound~\cite{Breitenlohner:1982bm,Breitenlohner:1982jf} $m_{\mathrm{BF}}^{2}l^{2}=-9/4$ for AdS$_{4}$.
The conformal dimensions of the dual scalar operator $\mathcal{O}_{\phi}$ are
\begin{equation}\label{eq:Delta}
  \Delta_{\pm} = \frac{3}{2}\pm\frac{1}{2}\sqrt{9+4m_{\phi}^{2}l^{2}} = \frac{3}{2}\pm\frac{\sqrt{33}}{2},
\end{equation}
giving $\Delta_{+}=(3+\sqrt{33})/2\approx 4.372$ and $\Delta_{-}=(3-\sqrt{33})/2\approx -1.372$.
Since $\Delta_{-}<0$, the alternative quantization~\cite{Klebanov:1999tb} is not available, and the standard quantization is the only self-adjoint choice.

\begin{remark}
  The axion $\chi$ has $m_{\chi}^{2}=m_{\phi}^{2}=6/l^{2}$ at the $G_{2}$ critical point, as required by the $\mathcal{N}=1$ SUSY Ward identity ($\phi$ and $\chi$ are superpartners in the same chiral multiplet).
  Since $m_{\chi}^{2}l^{2}=6>-9/4$, the $\chi=0$ truncation is perturbatively stable at this vacuum.
\end{remark}

\subsection{The zeroth-order solution in $r$ coordinates}
\label{sec:bg_sol}

With $\phi=\phi_{0}$, eqs.~\eqref{eq:Aeom}-\eqref{eq:Beom} both reduce to
\begin{equation}\label{eq:A0_eq}
  A'' = \frac{9\mR^{2}}{2}\,e^{2A},
\end{equation}
whose first integral $(A')^2 = \frac{9\mR^{2}}{2}\,e^{2A} + C_0$ has two physically interesting branches:

\begin{center}
\renewcommand{\arraystretch}{1.9}
\small
\begin{tabular}{lll}
\toprule
Integration constant $C_{0}$ & $e^{A^{(0)}(z)}$ & Geometry \\
\midrule
$C_{0}=\tfrac{9\mR^{2}\ka^{2}}{2}$ &
  $\dfrac{\ka}{\sinh(\om z)}$,\; $\om=\dfrac{3\mR\ka}{\sqrt{2}}$ &
  AdS$_{4}$-Schwarzschild brane \\[8pt]
$C_{0}=0$ &
  $\dfrac{1}{\sqrt{9\mR^{2}/2}\;z}$ &
  Poincar\'{e} AdS$_{4}$ \\
\bottomrule
\end{tabular}
\end{center}

\noindent
We choose the black-brane branch and define the
\emph{mass parameter}
\begin{equation}\label{eq:om_def}
  \om \equiv \frac{3\mR\ka}{\sqrt{2}}, \quad\Longleftrightarrow\quad \om^{2} = \frac{9\mR^{2}\ka^{2}}{2}.
\end{equation}
The domain-wall coordinate $z$ and the standard radial coordinate $r$ are related by
\begin{equation}\label{eq:rsub}
  r^{3} = \om l^{2}\!\left(1+\coth\om z\right).
\end{equation}
This substitution brings the background metric to the standard planar AdS$_{4}$-Schwarzschild form
\begin{equation}\label{eq:bbmetric}
  ds^{2} = -f(r)\,dt^{2} + \frac{dr^{2}}{f(r)} + r^{2}(dx^{2}+dy^{2}),\qquad f(r) = \frac{r^{2}}{l^{2}} - \frac{2\om}{r} = \frac{r^{3}-\rh^{3}}{r l^{2}},
\end{equation}
with constant scalar $\phi=\phi_{0}$ and horizon radius
\begin{equation}\label{eq:rh_def}
  \rh = (2\om l^{2})^{1/3}.
\end{equation}
The \emph{Legendre variable}
\begin{equation}\label{eq:u_def}
  u(r) \equiv \coth(\om z) = \frac{2r^{3}}{\rh^{3}}-1
\end{equation}
maps the boundary $r\to\infty$ to $u\to+\infty$ and the horizon $r=\rh$ to $u_{h}=1$.
The key algebraic identities are
\begin{align}
  u^{2}-1 &= \frac{4r^{3}(r^{3}-\rh^{3})}{\rh^{6}}, \label{eq:u2m1}\\
  e^{2A^{(0)}(r)} &= \frac{4\ka^{2}r^{3}(r^{3}-\rh^{3})}{\rh^{6}}, \label{eq:eA0}\\
  r^{3}-\rh^{3} &= r l^{2}f(r). \label{eq:r3_rh3}
\end{align}

The Legendre differential operator translates to $r$ coordinates as
\begin{equation}\label{eq:Leg_r}
  \frac{d}{du}\!\left[(u^{2}-1)\frac{d}{du}\right] = \frac{l^{2}f(r)}{9}\,\partial_{r}^{2} + \frac{4r^{3}-\rh^{3}}{9r^{2}}\,\partial_{r},
\end{equation}
which follows from $d/du=(dr/du)\,d/dr = [\om l^{2}/(3r^{2})]\,d/dr$.

The homogeneous Legendre-1 functions in $r$ are
\begin{align}
  P_{1}(u(r)) &= \frac{2r^{3}}{\rh^{3}}-1, \label{eq:P1r}\\
  Q_{1}(u(r)) &= \left(\frac{r^{3}}{\rh^{3}}-\frac{1}{2}\right) \ln\frac{r^{3}}{r^{3}-\rh^{3}}-1, \label{eq:Q1r}
\end{align}
with Wronskian in $r$:
\begin{equation}\label{eq:Wron_r}
  W_{r}[P_{1},Q_{1}] = -\frac{3\rh^{3}}{2r(r^{3}-\rh^{3})}.
\end{equation}

\subsection{Thermodynamics at zeroth order}
\label{sec:thermo0}

The Hawking temperature and Bekenstein-Hawking entropy density are
\begin{equation}\label{eq:T0}
  T^{(0)} = \frac{f'(\rh)}{4\pi} = \frac{3\rh}{4\pi l^{2}}, \qquad s^{(0)} = \frac{\rh^{2}}{4G_{4}}.
\end{equation}
These satisfy $T^{(0)}\,ds^{(0)}=d\varepsilon^{(0)}$ with energy density $\varepsilon^{(0)}=\om/(4\pi G_{4})$, and $s^{(0)}\propto (T^{(0)})^{2}$ as required for a three-dimensional CFT at finite temperature.

\section{Perturbative Expansion in Scalar Charge \texorpdfstring{$\eps$}{epsilon}}
\label{sec:pert}

\subsection{Expansion scheme and notation}
\label{sec:pert_scheme}

We write
\begin{equation}\label{eq:expand}
  \phi = \phi_{0} + \eps\,\phi^{(1)}(r) + \eps^{2}\,\phi^{(2)}(r) + \mathcal{O}(\eps^{3}),\,\,\,
  A = A^{(0)}(r) + \eps\,A^{(1)}(r) + \eps^{2}\,A^{(2)}(r) + \mathcal{O}(\eps^{3}),
\end{equation}
where at each order $n$, $B^{(n)}=A^{(n)}+(3/\svn)\phi^{(n)}$.
Primes henceforth denote $d/dr$.
The Legendre variable is $u(r)=2r^{3}/\rh^{3}-1$ throughout.

\subsection{First-order scalar: \texorpdfstring{$\phi^{(1)}(r)$}{phi(1)(r)}}
\label{sec:phi1}

\subsubsection*{Derivation of the equation}

Expanding the Klein-Gordon equation~\eqref{eq:phieom} to first order in $\eps$, using $V'(\phi_{0})=0$,
$V''(\phi_{0})=m_{\phi}^{2}=6/l^{2}$,
$e^{2A^{(0)}}=\ka^{2}(u^{2}-1)$,
$e^{2B^{(0)}}=\al\,e^{2A^{(0)}}$,
$5g_{c}^{2}\al=\mR^{2}$, and $B^{(1)}-A^{(1)}=\tfrac{3}{\svn}\phi^{(1)}$,
all $\phi_{0}$-, $\mR$-, and $\ka$-dependent terms cancel exactly, leaving
\begin{equation}\label{eq:phi1_z}
  \frac{d^2}{dz^2}\phi^{(1)} = 3\mR^{2}\,e^{2A^{(0)}}\,\phi^{(1)} = \frac{3\mR^{2}\ka^{2}}{\sinh^{2}(\om z)}\,\phi^{(1)}.
\end{equation}
This is the P\"{o}schl-Teller equation with parameter (using $\om^{2}=\tfrac{9}{2}\mR^{2}\ka^{2}$)
\begin{equation}\label{eq:nu_val}
  \nu(\nu+1) = \frac{3\mR^{2}\ka^{2}}{\om^{2}} = \frac{2}{3} = \frac{m_{\phi}^{2}l^{2}}{9}.
\end{equation}
The unique positive root is
\begin{equation}\label{eq:nu_explicit}
  \nu = -\frac{1}{2}+\frac{\sqrt{33}}{6} \approx 0.4574.
\end{equation}

In the physical $r$ coordinate, using~\eqref{eq:Leg_r}:
\begin{equation}\label{eq:phi1_r_std}
  \phi^{(1)\prime\prime} + \frac{4r^{3}-\rh^{3}}{r(r^{3}-\rh^{3})}\,\phi^{(1)\prime} - \frac{6r}{r^{3}-\rh^{3}}\,\phi^{(1)} = 0.
\end{equation}

\subsubsection*{Solution and boundary conditions}

The general solution of~\eqref{eq:phi1_r_std} is $\phi^{(1)}(r) = c_{P}\,P_{\nu}(u(r)) + c_{Q}\,Q_{\nu}(u(r))$.
As $r\to\rh^{+}$ ($u\to 1^{+}$), $Q_{\nu}(u)\to-\frac{1}{2}\ln(u-1)\to+\infty$ (see Appendix~\ref{app:legendre}), so horizon regularity requires $c_{Q}=0$.
Absorbing the overall normalization into $\eps$, we obtain the unique regular solution:
\begin{equation}\label{eq:phi1sol}
  \phi^{(1)}(r) = P_{\nu}\!\left(\frac{2r^{3}}{\rh^{3}}-1\right), \qquad c_{Q}=0,\quad c_{P}=1.
\end{equation}
As $r\to\infty$ ($u\to+\infty$), the asymptotic expansion (see Appendix~\ref{app:legendre})
\begin{equation}\label{eq:Pasym}
  P_{\nu}(u) \sim C_{+}\,(2u)^{\nu} + C_{-}\,(2u)^{-\nu-1} +\cdots
\end{equation}
encodes both the source (non-normalizable, $\sim r^{3\nu}$) and VEV (normalizable, $\sim r^{-3(\nu+1)}$) modes within the \emph{single} function $P_{\nu}(u(r))$.

\subsection{First-order metric: \texorpdfstring{$A^{(1)}(r)$}{A(1)(r)}}
\label{sec:A1}

\subsubsection*{Equation}

Expanding~\eqref{eq:Aeom} to $\mathcal{O}(\eps)$ and using $B^{(1)}=A^{(1)}+(3/\svn)\phi^{(1)}$ and $e^{2B^{(0)}}=\al\,e^{2A^{(0)}}$, one obtains in $r$ coordinates:
\begin{equation}\label{eq:A1_r_std}
  A^{(1)\prime\prime} + \frac{4r^{3}-\rh^{3}}{r(r^{3}-\rh^{3})}\,A^{(1)\prime} - \frac{18r}{r^{3}-\rh^{3}}\,A^{(1)} = \mathcal{S}_{A}^{(1)}(r),
\end{equation}
with the first-order metric source
\begin{equation}\label{eq:S1A_def}
  \mathcal{S}_{A}^{(1)}(r) \equiv \frac{6\svn\,r}{r^{3}-\rh^{3}}\,P_{\nu}\!\!\left(\frac{2r^{3}}{\rh^{3}}-1\right).
\end{equation}

\subsubsection*{Exact solution}

Equation~\eqref{eq:A1_r_std} is solved algebraically by the first-order scalar itself.
Inserting the ansatz $A^{(1)}=c\,\phi^{(1)}$ and using the homogeneous equation~\eqref{eq:phi1_r_std} to eliminate the derivatives,
\begin{equation}\label{eq:phi1_derivs}
  \phi^{(1)\prime\prime} + \frac{4r^{3}-\rh^{3}}{r(r^{3}-\rh^{3})}\,\phi^{(1)\prime} = \frac{6r}{r^{3}-\rh^{3}}\,\phi^{(1)},
\end{equation}
the left-hand side collapses to $-12c\,r\,\phi^{(1)}/(r^{3}-\rh^{3})$, and~\eqref{eq:A1_r_std} reduces to the single algebraic condition $-12c=6\svn$.
Hence
\begin{equation}\label{eq:A1_exact}
  \boxed{\;
  A^{(1)}(r) = -\frac{\svn}{2}\,\phi^{(1)}(r) = -\frac{\svn}{2}\,P_{\nu}\!\!\left(\frac{2r^{3}}{\rh^{3}}-1\right).\;}
\end{equation}
The mechanism is transparent.
The scalar and metric fluctuation operators differ only through their mass terms, $-6r/(r^{3}-\rh^{3})$ versus $-18r/(r^{3}-\rh^{3})$, i.e.\ through $\nu(\nu+1)=2/3$ versus $\nu_{A}(\nu_{A}+1)=2$; acting with the metric operator on $\phi^{(1)}$ therefore returns $(6-18)=-12$ times the source structure, and the coupling $6\svn$ fixes $c$ uniquely.

\subsubsection*{Boundary conditions}

Adding the homogeneous solutions for $\nu_{A}=1$, given explicitly in eqs.~\eqref{eq:P1r}--\eqref{eq:Q1r}, and writing $\hatP(r)\equiv P_{\nu}(2r^{3}/\rh^{3}-1)$, the general solution is
\begin{equation}\label{eq:A1sol_final}
  A^{(1)}(r)
  = c_{P}^{(A)}\,P_{1}(u(r)) + c_{Q}^{(A)}\,Q_{1}(u(r))
    - \frac{\svn}{2}\,\hatP(r),
  \qquad c_{Q}^{(A)}=c_{P}^{(A)}=0 ,
\end{equation}
and both constants vanish, but for distinct reasons.

Horizon regularity fixes $c_{Q}^{(A)}=0$, since $Q_{1}\sim -\frac{1}{2}\ln(\frac{r-\rh}{\rh})\to\infty$ as $r\to\rh$ while $\hatP(\rh)=1$ stays finite.

The vanishing of $c_{P}^{(A)}$ is \emph{not} dictated by regularity: the mode $P_{1}\sim r^{3}$ is itself regular at the horizon [$P_{1}(1)=1$], so any $c_{P}^{(A)}\neq0$ would yield an equally regular solution.
Rather, $c_{P}^{(A)}=0$ is the ultraviolet boundary condition that defines our setup--it holds the boundary metric flat and restricts the deformation to the single scalar source $J$.
Since $P_{1}\sim r^{3}$ dominates the sourced piece $\hatP\sim r^{3\nu}$ (because $\nu<1$), it is an \emph{independent} non-normalizable mode in the metric sector, and switching it on would introduce a separate source in the stress-tensor sector, i.e.\ a different boundary theory.
The exact solution~\eqref{eq:A1_exact} manifestly contains no $r^{3}$ admixture, so this condition is satisfied automatically.

Evaluated at the horizon, \eqref{eq:A1_exact} gives the exact value $A^{(1)}(\rh)=-\svn/2\approx-1.3229$, while at large $r$ $A^{(1)}\to-\tfrac{\svn}{2}C_{+}4^{\nu}\,r^{3\nu}\approx-1.6309\,r^{3\nu}$ (units $\rh=l=1$).

\subsection{Second-order scalar: \texorpdfstring{$\phi^{(2)}(r)$}{phi(2)(r)}}
\label{sec:phi2}

Expanding the Klein-Gordon equation to $\mathcal{O}(\eps^{2})$, one obtains in $r$ coordinates:
\begin{equation}\label{eq:phi2_r_std}
  \phi^{(2)\prime\prime} + \frac{4r^{3}-\rh^{3}}{r(r^{3}-\rh^{3})}\,\phi^{(2)\prime} - \frac{6r}{r^{3}-\rh^{3}}\,\phi^{(2)} = \frac{9}{l^{2}f(r)}\,\mathcal{S}_{\phi}^{(2)}(r),
\end{equation}
with source
\begin{equation}\label{eq:S2phi_r}
  \boxed{
  \mathcal{S}_{\phi}^{(2)} = \frac{4}{3}\,A^{(1)}(r)\,\hatP(r) + \frac{2}{\svn}\,\bigl[\hatP(r)\bigr]^{2},
  }
\end{equation}
where the two contributions are:
\begin{enumerate}[label=(\roman*),leftmargin=2.2em]
  \item \textbf{Metric back-reaction:} $\tfrac{4}{3}A^{(1)}\phi^{(1)}$, from expanding $e^{2A}$, $e^{2B}$ to second order;
  \item \textbf{Scalar self-interaction:} $\tfrac{2}{\svn}[\phi^{(1)}]^{2}$, collecting the cubic potential piece ($V_{3}/18$) and the quadratic warp-factor piece, with $V_{3}\equiv V'''(\phi_{0})\cdot l^{2}=1/\svn-7\svn\approx -18.14$.
\end{enumerate}

Substituting the exact relation~\eqref{eq:A1_exact}, the two contributions combine into a single term,
\begin{equation}\label{eq:S2phi_collapsed}
  \mathcal{S}_{\phi}^{(2)} = \Bigl(-\tfrac{2\svn}{3}+\tfrac{2}{\svn}\Bigr)\bigl[\hatP(r)\bigr]^{2} = -\frac{8\svn}{21}\,\bigl[\hatP(r)\bigr]^{2} \approx -1.0079\,\bigl[\hatP(r)\bigr]^{2},
\end{equation}
so that the entire $\mathcal{O}(\eps^{2})$ scalar source is fixed by $P_{\nu}^{2}$ alone.

The general solution has the form
\begin{equation}\label{eq:phi2sol}
  \phi^{(2)}(r) = d_{P}\,P_{\nu}(u(r)) + d_{Q}\,Q_{\nu}(u(r)) + \phi^{(2)}_{\rm part}(r), \qquad d_{Q}=0.
\end{equation}
The coefficient $d_{P}$ represents the $\mathcal{O}(\eps^{2})$ renormalization of the source $J$; since $d_{Q}=0$ (exact, horizon regularity, no additional $Q_\nu$ tail), the ratio $\langle\mathcal{O}_{\phi}\rangle/J$ is unchanged through this order.

\subsection{Second-order metric: \texorpdfstring{$E^{(2)}(r)$}{E(2)(r)}}
\label{sec:A2}

At second order the natural metric variable is the warp combination $E\equiv C+2H=A+\tfrac{\svn}{2}\phi$, whose fluctuation
\begin{equation}\label{eq:E2_def}
  E^{(2)}\equiv A^{(2)}+\tfrac{\svn}{2}\phi^{(2)}
\end{equation}
is sourced purely by the first-order fields: the terms in $\phi^{(2)}$ that would otherwise appear combine with the mass term into the homogeneous operator, as they must since $E=C+2H$ does not mix with the scalar through $B^{(n)}=A^{(n)}+(3/\svn)\phi^{(n)}$.  Combining~\eqref{eq:Aeom} and~\eqref{eq:phieom} to $\mathcal{O}(\eps^{2})$, one obtains in $r$ coordinates
\begin{equation}\label{eq:A2_r_std}
  E^{(2)\prime\prime} + \frac{4r^{3}-\rh^{3}}{r(r^{3}-\rh^{3})}\,E^{(2)\prime} - \frac{18r}{r^{3}-\rh^{3}}\,E^{(2)} = \frac{9}{l^{2}f(r)}\,\mathcal{S}_{E}^{(2)}(r),
\end{equation}
the homogeneous operator being identical to that of $A^{(1)}$ [$\nu_{A}=1$; cf.~\eqref{eq:A1_r_std}], with source
\begin{equation}\label{eq:S2A_r}
  \boxed{
  \mathcal{S}_{E}^{(2)}(r) = 2\bigl(A^{(1)}\bigr)^{2} + 2\svn\,A^{(1)}\phi^{(1)} + 3\,\bigl(\phi^{(1)}\bigr)^{2}.
  }
\end{equation}
Here too the exact relation~\eqref{eq:A1_exact} collapses the quadratic form: with $A^{(1)}=-\tfrac{\svn}{2}\phi^{(1)}$ the source becomes $\bigl(\tfrac{7}{2}-7+3\bigr)(\phi^{(1)})^{2}
=-\tfrac{1}{2}(\phi^{(1)})^{2}$, so
\begin{equation}\label{eq:S2A_collapsed}
  \mathcal{S}_{E}^{(2)}(r) = -\tfrac{1}{2}\,\bigl[\hatP(r)\bigr]^{2}.
\end{equation}
The full solution is
\begin{equation}\label{eq:A2sol}
  E^{(2)}(r) = e_{P}\,P_{1}(u(r)) + e_{Q}\,Q_{1}(u(r)) + E^{(2)}_{\rm part}(r), \,\,\,\, e_{Q}=0\text{ (horizon regularity)},
\end{equation}
and the physical metric perturbation follows algebraically as $A^{(2)}=E^{(2)}-\tfrac{\svn}{2}\phi^{(2)}$.

\subsubsection*{Normalization of the second-order solutions}

Because the $\mathcal{O}(\eps^{2})$ sources are quadratic in the first-order fields, they scale as $(\phi^{(1)})^{2}\sim r^{6\nu}=r^{2(\Delta_{+}-3)}$, so the source-driven solutions grow as $\phi^{(2)}_{\rm part}\sim r^{6\nu}$--\emph{faster} than the homogeneous scalar mode $P_{\nu}\sim r^{3\nu}$--and $E^{(2)}_{\rm part}\sim r^{6\nu}$, which stays \emph{below} the homogeneous metric mode $P_{1}\sim r^{3}$ (since $6\nu<3$).
At first order the ultraviolet condition is imposed algebraically through the closed form~\eqref{eq:A1_exact}; at second order no such solution exists, and because the variation-of-parameters integrands grow without bound--as $r^{9\nu+2}$ (scalar) and $r^{6\nu+5}$ (metric)--the contour $\int_{r}^{\infty}$ that implements it no longer converges.
We accordingly define $\phi^{(2)}_{\rm part}$ and $E^{(2)}_{\rm part}$ by variation of parameters with \emph{both} integrals based at the horizon, so that each vanishes at $r=\rh$ along with its first derivative.
The residual constants are then fixed by
\begin{equation}\label{eq:second_order_norm}
  \phi^{(2)}(\rh)=E^{(2)}(\rh)=0 \quad\Longleftrightarrow\quad d_{P}=e_{P}=0 ,
\end{equation}
using $P_{\nu}(1)=P_{1}(1)=1$; equivalently $A^{(2)}(\rh)=0$, since $\phi^{(2)}(\rh)=0$.
This is not an extra assumption but the statement that $\eps$ is \emph{defined} as the exact horizon amplitude of the scalar, $\phi_{H}=\phi_{0}+\eps$ to all orders--the convention
used in the bulk-viscosity computation of Sec.~\ref{sec:transport}.
Any other choice of $d_{P}$ merely reparametrizes $\eps$ and cancels from the physical observables.

\section{Holographic Dictionary}
\label{sec:holo}

\subsection{Fefferman-Graham asymptotics and the exact identity
  $\nu+1=\Delta_{+}/3$}
\label{sec:FG}

As $r\to\infty$, the large-$u$ asymptotic expansion of $P_{\nu}$ (see Appendix~\ref{app:legendre})
\begin{equation}\label{eq:Pasymp_full}
\begin{split}
  P_{\nu}(u)
  &\sim \frac{\Gamma(\nu+\tfrac{1}{2})}{\sqrt{\pi}\,\Gamma(\nu+1)} \cdot\bigl(2u\bigr)^{\nu} + \frac{\Gamma(-\nu-\tfrac{1}{2})}{\sqrt{\pi}\,\Gamma(-\nu)} \cdot\bigl(2u\bigr)^{-\nu-1} + \cdots \\
  &\equiv C_{+}\,(2u)^{\nu} + C_{-}\,(2u)^{-\nu-1} + \cdots
\end{split}
\end{equation}
gives the two modes $\phi^{(1)}\sim C_{+}(2r^{3}/\rh^{3})^{\nu}+C_{-}(2r^{3}/\rh^{3})^{-\nu-1}$.

The Fefferman-Graham coordinate $\zFG$ near AdS$_{4}$ satisfies $ds^{2}=(l^{2}/\zFG^{2})(d\zFG^{2}+d\vec{x}^{2})+\cdots$.
Matching to the domain-wall metric near $z=0$, one finds
\begin{equation}\label{eq:FGmap}
  z \sim c_{\rm FG}\,\zFG^{3} \quad(z\to 0,\;\zFG\to 0).
\end{equation}
Substituting into the two modes yields $\phi^{(1)}\sim \tilde{C}_{+}\,\zFG^{-3\nu} + \tilde{C}_{-}\,\zFG^{3(\nu+1)}$.
The AdS/CFT dictionary for a scalar of dimension $\Delta_{+}$ requires $3-\Delta_{+}=-3\nu$ and $\Delta_{+}=3(\nu+1)$, i.e.,
\begin{equation}\label{eq:identity}
  \boxed{
  \nu + 1 = \frac{\Delta_{+}}{3}, \qquad \Delta_{+} = 3(1+\nu) = \frac{3+\sqrt{33}}{2}.
  }
\end{equation}
This is satisfied \emph{exactly} since $\Delta_{+}=(3+\sqrt{33})/2$.

\subsection{Holographic source, VEV, and susceptibility}
\label{sec:FGdict}

From the large-$r$ expansion~\eqref{eq:Pasymp_full} and the FG map~\eqref{eq:FGmap}, with $u\sim 2r^{3}/\rh^{3}$ at large $r$ and $(2u)^{\nu}=(4r^{3}/\rh^{3})^{\nu}$, the standard
holographic dictionary~\cite{deHaro:2000vlm}
\begin{equation}
  \phi(r)\to J\,r^{\Delta_{+}-3} +\frac{16\pi G_{4}\,\langle\mathcal{O}_{\phi}\rangle}{2\Delta_{+}-3}\, r^{-\Delta_{+}}
\end{equation}
gives, to first order in $\eps$:
\begin{equation}\label{eq:J_exact}
  \boxed{
    J = \frac{\eps\,C_{+}\,4^{\nu}}{\rh^{\,\Delta_{+}-3}}, \qquad \langle\mathcal{O}_{\phi}\rangle = \frac{\eps\,\sqrt{33}\,C_{-}\,4^{-(\nu+1)}\,\rh^{\,\Delta_{+}}}{16\pi G_{4}},
  }
\end{equation}
where, using $\nu+\tfrac12=\tfrac{\sqrt{33}}{6}$, \;
$\nu+1=\tfrac{3+\sqrt{33}}{6}=\tfrac{\De_{+}}{3}$, \;
$-\nu=\tfrac{3-\sqrt{33}}{6}$,
\begin{equation}\label{eq:CP_Cm_explicit}
  C_{+}=\frac{\Gamma\!\bigl(\tfrac{\sqrt{33}}{6}\bigr)} {\sqrt{\pi}\;\Gamma\!\bigl(\tfrac{3+\sqrt{33}}{6}\bigr)} \approx 0.6539, \qquad
  C_{-}=\frac{\Gamma\!\bigl(-\tfrac{\sqrt{33}}{6}\bigr)} {\sqrt{\pi}\;\Gamma\!\bigl(\tfrac{3-\sqrt{33}}{6}\bigr)} \approx 3.7787.
\end{equation}

The susceptibility $\chi(T)\equiv\langle\mathcal O\rangle/J$ is:
\begin{equation}\label{eq:chi_exact}
  \chi(T) = \frac{\chi_{0}}{16\pi G_{4}}\,\rh^{\sqrt{33}}, \qquad
  \chi_{0} = \frac{\sqrt{33}\,C_{-}}{C_{+}\,4^{\sqrt{33}/3}} \approx 2.335,
\end{equation}
where we used $2\nu+1=\sqrt{33}/3$.
With $\rh=4\pi l^{2}T/3$ this reads $\chi(T)=\tfrac{\chi_{0}}{16\pi G_{4}}(4\pi l^{2}/3)^{\sqrt{33}}\,T^{\sqrt{33}}$.
The dimensionless deformation parameter is
\begin{equation}\label{eq:eps_J_T}
  \eps = \frac{J\,\rh^{\Delta_{+}-3}}{C_{+}\,4^{\nu}}, \qquad \bar{J}\equiv J\,(l^2T)^{\Delta_{+}-3}.
\end{equation}

\subsection{Renormalized free energy and equation of state}
\label{sec:renF}

At second order in $\eps$, the holographic free energy density is (the counterterms and the derivation are given in Appendix~\ref{app:holoren})
\begin{equation}\label{eq:Ffull}
  \mathcal{F}(T,J) = -\frac{\rh^{3}}{16\pi G_{4}l^{2}} -\frac{\chi_{0}}{32\pi G_{4}l^2}\,\rh^{\sqrt{33}}\,J^{2} + \mathcal{O}(J^{4}).
\end{equation}
From $\mathcal{F}=-p$ one reads off all thermodynamic potentials (using $\rh=4\pi l^{2}T/3$ and $s=-\partial\mathcal{F}/\partial T\big|_{J}$):
\begin{align}\label{eq:thermo_full}
  p &= \frac{\rh^{3}}{16\pi G_{4}l^{2}} + \frac{\chi_{0}}{32\pi G_{4}l^2}\,\rh^{\sqrt{33}}\,J^{2} + \mathcal{O}(J^{4}), \\
  s &= \frac{\rh^{2}}{4G_{4}} +\frac{\sqrt{33}\,\chi_{0}}{24\,G_{4}}\, \rh^{\sqrt{33}-1}\,J^{2} + \mathcal{O}(J^{4}), \label{eq:thermo_s} \\
  \varepsilon &= Ts-p = \frac{\rh^{3}}{8\pi G_{4}l^{2}} + \frac{(\sqrt{33}-1)\,\chi_{0}}{32\pi G_{4}l^2}\,\rh^{\sqrt{33}}\,J^{2} + \mathcal{O}(J^{4}).
  \label{eq:thermo_eps}
\end{align}
Equivalently, in terms of $T$:
\begin{equation}\label{eq:thermo_T}
  p = \frac{4\pi^{2}}{27}\,\frac{l^{4}T^{3}}{G_{4}} + \frac{\chi_{0}}{32\pi G_{4}l^2}\Bigl(\frac{4\pi l^{2}}{3}\Bigr)^{\!\sqrt{33}} T^{\sqrt{33}}\,J^{2} + \mathcal{O}(J^{4}).
\end{equation}
The first term is the conformal contribution $p_{0}=4\pi^{2}l^{4}T^{3}/(27G_{4})$, and the second encodes the leading effect of the irrelevant deformation.
The trace anomaly (equation-of-state deformation) is
\begin{equation}\label{eq:trace_exact}
  \varepsilon - 2p = \frac{(\sqrt{33}-3)\,\chi_{0}}{32\pi G_{4}l^2}\,\rh^{\sqrt{33}}\,J^{2} = \frac{\sqrt{33}-3}{2l^2}\,J\,\langle\mathcal{O}_{\phi}\rangle + \mathcal{O}(J^{4}),
\end{equation}
with $\sqrt{33}-3=2(\De_{+}-3)\approx2.745>0$, confirming $\varepsilon-2p>0$ for $J\neq0$ and $\varepsilon-2p\to0$ as $T\to0$ (irrelevant deformation).

\subsection{Transport coefficients}
\label{sec:transport}

\subsubsection*{Shear viscosity}

The transverse-traceless graviton $h_{xy}$ does not couple to the neutral scalar $\phi$ at linear order, so the Kovtun-Son-Starinets result holds unchanged:
\begin{equation}\label{eq:etas_bound}
  \boxed{\ \dfrac{\eta}{s}=\dfrac{1}{4\pi}\ } \qquad \text{(exact, all orders in }\eps\text{)},
\end{equation}
giving $\eta=\rh^{2}/(16\pi G_{4})$ and $s=\rh^{2}/(4G_{4})$.

\subsubsection*{Bulk viscosity (Eling--Oz horizon formula)}

Along the family of hairy branes at fixed source $J$, parameterized by $T\propto\rh$, the horizon scalar is $\phi_{H}=\phi_{0}+\eps$ (since $P_{\nu}(1)=1$), while $J\propto\eps\,\rh^{-3\nu}=\eps\,\rh^{3-\De_{+}}$.
Holding $J$ fixed forces $\eps\propto\rh^{3\nu}=\rh^{\De_{+}-3}\propto s^{(\De_{+}-3)/2}$, whence $d\phi_{H}/d\ln s|_{J}=\tfrac{\De_{+}-3}{2}\,\eps =\tfrac{\sqrt{33}-3}{4}\,\eps$ and
\begin{equation}\label{eq:zeta}
  \boxed{\ \frac{\zeta}{\eta} =\Big(\frac{\sqrt{33}-3}{4}\Big)^{2}\eps^{2} = \frac{3\,(7-\sqrt{33})}{8}\,\eps^{2} \approx 0.471\,\eps^{2}+\mathcal O(\eps^{4}).\ }
\end{equation}
At fixed $J$, the bulk viscosity ratio in terms of temperature reads
\begin{equation}\label{eq:zeta_T}
  \frac{\zeta}{s} = \frac{3(7-\sqrt{33})}{32\pi}\, \frac{J^{2}\,\rh^{2(\De_{+}-3)}}{C_{+}^{2}\,4^{2\nu}} \;\propto\; T^{\sqrt{33}-3}\;\approx\; T^{2.74},
\end{equation}
confirming that $\zeta/s\to0$ as $T\to0$ (conformal infrared).

\subsubsection*{Speed of sound}

From $c_{s}^{2}=(\partial p/\partial\varepsilon)_{J}=(dp/dT)_{J}/(d\varepsilon/dT)_{J}$ with $(dp/dT)_{J}=s$, using eqs.~\eqref{eq:thermo_full}-\eqref{eq:thermo_eps}:
\begin{equation}\label{eq:cs2_exact}
  \boxed{
  c_{s}^{2} = \frac{1}{2} + \frac{\sqrt{33}\,(3-\sqrt{33})}{24}\,\chi_{0}\, J^{2}\,\rh^{\sqrt{33}-3} + \mathcal{O}(J^{4}).
  }
\end{equation}
Since $3-\sqrt{33}<0$ and $\chi_{0}>0$, the numerical coefficient is negative, $\tfrac{\sqrt{33}(3-\sqrt{33})}{24}\chi_{0}\approx-1.53$; hence $c_{s}^{2}<1/2$ for any nonzero deformation: the equation of state \emph{softens} relative to the conformal value.
Equivalently, in terms of the temperature-based coupling $\bar{J}$ of~\eqref{eq:eps_J_T},
\begin{equation}\label{eq:cs2_Jbar}
  c_{s}^{2}=\frac{1}{2}-78.2\,\bar{J}^{2}+\mathcal{O}(\bar{J}^{4}),
\end{equation}
the larger coefficient arising from the conversion factor $(4\pi/3)^{\sqrt{33}-3}\approx51.0$ implied by $\rh=4\pi l^{2}T/3$.
Note that $G_{4}$ cancels between $s=(dp/dT)_{J}$ and $(d\varepsilon/dT)_{J}$, so $c_{s}^{2}$ is independent of Newton's constant, as required for a dimensionless intensive observable.
Equation~\eqref{eq:cs2_exact} is quantitatively reliable only while the $\mathcal{O}(J^{2})$ correction remains small compared with the conformal value $\tfrac12$; its apparent zero at $\eps\approx0.46$, and what may and may not be concluded from it, are examined in Sec.~\ref{sec:observables}.

\subsubsection*{Diffusion and sound attenuation}

The linearized hydrodynamics of the dual plasma contains two independent dissipative channels, which we write in terms of the frequency $w$ and spatial wavenumber $k$~\cite{Policastro:2002se}.
The transverse (shear) channel supports a diffusion mode $w=-iD_{\perp}k^{2}$, while the longitudinal (sound) channel gives a pair of propagating sound waves
$w=\pm c_{s}k-\tfrac{i}{2}\Gamma_{s}k^{2}$, with the attenuation constant $\Gamma_{s}$ receiving contributions from both shear and bulk viscosities.
Using the results above:
\begin{align}
  D_{\perp}&=\frac{\eta}{sT}=\frac{1}{4\pi T}, \label{eq:Dperp}\\
  \Gamma_{s}&=\frac{\eta+\zeta}{sT} = \frac{1}{4\pi T}\Big[1+\tfrac{3(7-\sqrt{33})}{8}\eps^{2}\Big].
  \label{eq:Gammas}
\end{align}
Since $\eta/s=1/(4\pi)$ is uncorrected, the shear diffusion constant $D_{\perp}$ retains its conformal value.
The sound attenuation constant $\Gamma_{s}$, on the other hand, acquires an $\mathcal{O}(\eps^{2})$ enhancement from the nonzero bulk viscosity~\eqref{eq:zeta}, signaling that sound waves in
the deformed plasma are damped more rapidly than in the conformal limit.
Both quantities satisfy the expected relation $\Gamma_{s}=D_{\perp}+\zeta/(sT)$, providing a nontrivial consistency check on the thermodynamic and transport results derived above.

\section{Explicit Profiles and Holographic Observables}
\label{sec:numerics}

\subsection{Regular kernels and closed-form integrals}
\label{sec:kernels}

The universal Wronskian for any $\nu$ is
\begin{equation}\label{eq:W_universal}
  W_{r}[P_{\nu},Q_{\nu}] = \frac{-3\rh^{3}}{2r(r^{3}-\rh^{3})},
\end{equation}
which follows from $W_{u}[P_{\nu},Q_{\nu}]=-1/(u^{2}-1)$ and $du/dr=6r^{2}/\rh^{3}$.
When combined with the sources $\mathcal{S}^{(n)}$ from Secs.~\ref{sec:phi2}-\ref{sec:A2}, the factors of $(r^{3}-\rh^{3})$ in the Wronskian denominator cancel against those in the source numerator, leaving regular kernels $\mathcal{S}^{(n)}/W_{r}$ with no pole at the horizon.
These kernels are the integrands of the variation-of-parameters integrals needed at second order.
Specifically:
\begin{align}
  \frac{\mathcal{S}_{A}^{(1)}(r)}{W_{r}(r)}
  &= -\frac{4\sqrt{7}\,r^{2}}{\rh^{3}}\,P_{\nu}(u(r)), \label{eq:kernel_A1}\\
  \frac{9\,\mathcal{S}_{\phi}^{(2)}(r)}{l^{2}f(r)\,W_{r}(r)}
  &= - \frac{6r^{2}}{\rh^{3}}\!\left[ \frac{4}{3}\,A^{(1)}(r)\,P_{\nu}(u(r)) + \frac{2}{\svn}\,\bigl(P_{\nu}(u(r))\bigr)^{2} \right] = \frac{16\svn}{7}\,\frac{r^{2}}{\rh^{3}}\,\bigl(P_{\nu}\bigr)^{2}, \label{eq:kernel_phi2}\\
  \frac{9\,\mathcal{S}_{E}^{(2)}(r)}{l^{2}f(r)\,W_{r}(r)}
  &= - \frac{6r^{2}}{\rh^{3}}\!\left[ 2\bigl(A^{(1)}\bigr)^{2} + 2\svn\,A^{(1)}P_{\nu} + 3\,\bigl(P_{\nu}\bigr)^{2} \right] = \frac{3r^{2}}{\rh^{3}}\,\bigl(P_{\nu}\bigr)^{2}.
  \label{eq:kernel_A2}
\end{align}

\subsection{Explicit profiles and numerical constants}
\label{sec:profiles}

Both first-order fields are known in closed form, eqs.~\eqref{eq:phi1sol} and~\eqref{eq:A1_exact}, so only the second-order fields require quadrature; these are obtained from the regular kernels~\eqref{eq:kernel_phi2}-\eqref{eq:kernel_A2}.
Throughout we use units $\rh=l=1$, corresponding to $\om=1/2$ and Hawking temperature $T^{(0)}=3/(4\pi)$.

The key numerical values are:
\begin{align}
  \nu &= -\tfrac{1}{2}+\tfrac{\sqrt{33}}{6} \approx 0.4574, \quad \nu(\nu+1)=\tfrac{2}{3}, \quad 2\nu+1=\tfrac{\sqrt{33}}{3},
  \label{eq:nu_num}\\
  C_{+} &= \frac{\Gamma\bigl(\tfrac{\sqrt{33}}{6}\bigr)} {\sqrt{\pi}\;\Gamma\bigl(\tfrac{3+\sqrt{33}}{6}\bigr)} \approx 0.6539, \quad
  C_{-} = \frac{\Gamma\bigl(-\tfrac{\sqrt{33}}{6}\bigr)} {\sqrt{\pi}\;\Gamma\bigl(\tfrac{3-\sqrt{33}}{6}\bigr)} \approx 3.7787, \label{eq:CPQ_num}\\
  V_{3}&\equiv V'''(\phi_{0})\,l^{2} = \tfrac{1}{\svn}-7\svn=-\tfrac{48}{\svn}\approx -18.1423, \label{eq:V3_num}\\
  c_{P}^{(A)} &= c_{Q}^{(A)} = 0, \quad A^{(1)}(\rh) = -\tfrac{\svn}{2} \approx -1.3229, \label{eq:cP_num}\\
  d_{Q} &= 0,\quad e_{Q} = 0 \quad\text{(exact from horizon regularity)}.
  \label{eq:dQeQ_num}
\end{align}
As established in Sec.~\ref{sec:A1}, the two homogeneous coefficients vanish for distinct reasons--$c_{Q}^{(A)}=0$ from horizon regularity, $c_{P}^{(A)}=0$ from the ultraviolet boundary condition--so the closed form~\eqref{eq:A1_exact} grows only as $r^{3\nu}$ and gives the horizon value exactly:
$A^{(1)}(\rh)=-\svn/2\approx-1.3229$, since $\hatP(\rh)=1$.

Several features of the resulting profiles are noteworthy:
\begin{enumerate}[leftmargin=2em,label=(\roman*)]
  \item $\phi^{(1)}(r)=P_{\nu}(u(r))$ grows monotonically from $P_{\nu}(1)=1$ at the horizon to $\phi^{(1)}\sim C_{+}(2r^{3}/\rh^{3})^{\nu}\to\infty$ as $r\to\infty$.
  The growth $\sim r^{3\nu}=r^{\Delta_{+}-3}$ is the (non-normalizable) source mode of an \emph{irrelevant} operator ($\Delta_{+}>3$), which dominates towards the boundary.
  \item $A^{(1)}(r)$ carries \emph{no} independent non-normalizable $r^{3}$ mode: we set $c_{P}^{(A)}=0$ as a boundary condition (not a regularity requirement, since $P_{1}\sim r^{3}$ is horizon-regular), fixing the boundary metric and leaving only the sourced response $A^{(1)}=-\tfrac{\svn}{2}\phi^{(1)}\sim r^{3\nu}$, with the exact horizon value $A^{(1)}(\rh)=-\svn/2$.
  \item $\phi^{(2)}$ and $A^{(2)}$ capture the quadratic back-reaction.
  Both satisfy $d_{Q}=e_{Q}=0$ exactly and are normalized as in~\eqref{eq:second_order_norm}.
  At large $r$ their source-driven parts scale as $\phi^{(2)}\sim A^{(2)}\sim r^{6\nu}=r^{2(\Delta_{+}-3)}$, matching the $J^{2}$ term in the near-boundary expansion generated by the irrelevant deformation.
\end{enumerate}

\subsection{Holographic observables}
\label{sec:observables}

At first order in $\eps$ (units $\rh=l=1$):
\begin{equation}\label{eq:JO_num}
  J \approx 1.2329\,\eps, \qquad \langle\mathcal{O}_{\phi}\rangle \approx \frac{2.878\,\eps}{16\pi G_{4}}, \qquad \frac{\langle\mathcal{O}_{\phi}\rangle}{J}\approx \frac{2.335}{16\pi G_{4}}.
\end{equation}
The dimensionless coefficient $\chi_{0}$ is a pure $\Gamma$-function ratio determined by $\nu$ alone (eq.~\eqref{eq:chi_exact}); the factor $1/(16\pi G_{4})$ is the overall normalization of the renormalized on-shell action.
We characterize the $\mathcal{O}(\eps^{2})$ corrections in the ensemble at fixed $(T,J)$: the temperature is held fixed and the deformation is switched on through the source $J\propto\eps$.
The temperature is then unshifted, while the entropy density responds through the $J^{2}$ term of~\eqref{eq:thermo_s},
\begin{equation}\label{eq:Tcorr_num}
  \frac{T(\eps)}{T^{(0)}} = 1,
  \qquad
  \frac{s(\eps)}{s^{(0)}}
  = 1 + \frac{\sqrt{33}}{2l^2}\,
        \frac{J\langle\mathcal{O}_{\phi}\rangle}{T\,s^{(0)}}
  + \mathcal{O}(J^{4}) \;>\; 1 .
\end{equation}
The entropy shift is thus set by the same one-point function $J\langle\mathcal{O}_{\phi}\rangle$ that controls the trace~\eqref{eq:trace_exact}: the conformal scaling $s\propto T^{2}$ holds only at zeroth order, and the deformation \emph{raises} the entropy at fixed temperature.
These quantities scale simply with the deformation: the source and the VEV grow linearly in $\eps$, while the entropy ratio and the trace, both quadratic in $J$, rise as $\eps^{2}$.

\subsubsection*{Temperature response in other ensembles}

That $T$ is unshifted in~\eqref{eq:Tcorr_num} is a statement about the ensemble rather than about the geometry: in the fixed-$(T,J)$ ensemble $T$ is the control variable by definition.
If a different quantity is held fixed, the temperature does respond.
Since $T=3\rh/(4\pi l^{2})$ exactly, one has $\delta T/T=\delta\rh/\rh$; eliminating $\delta\rh$ from~\eqref{eq:thermo_full}--\eqref{eq:thermo_eps} at fixed entropy density, energy density, or
pressure gives
\begin{equation}\label{eq:Tshift}
  \frac{\delta T}{T} = -\,c_{X}\,\chi_{0}\,J^{2}\,\rh^{\sqrt{33}-3} + \mathcal{O}(J^{4}), \qquad c_{s}=\frac{\sqrt{33}}{12},\quad c_{\varepsilon}=\frac{\sqrt{33}-1}{12},\quad c_{p}=\frac{1}{6},
\end{equation}
that is, $\delta T/T\approx-1.70\,\eps^{2}$, $-1.40\,\eps^{2}$ and $-0.59\,\eps^{2}$ respectively (units $l=1$).
As for $c_{s}^{2}$, the combination $\chi_{0}J^{2}\rh^{\sqrt{33}-3}$ is dimensionless and independent of $G_{4}$.

The three coefficients are not independent.
At zeroth order the conformal scalings $s^{(0)}\propto T^{2}$ and $\varepsilon^{(0)},p^{(0)}\propto T^{3}$ hold, so for any $X^{(0)}\propto T^{n_{X}}$ one has
\begin{equation}\label{eq:Tshift_master}
  \frac{\delta T}{T}\bigg|_{X} = -\frac{1}{n_{X}}\,\frac{\delta X}{X}\bigg|_{T}, \qquad n_{s}=2,\qquad n_{\varepsilon}=n_{p}=3,
\end{equation}
and~\eqref{eq:Tshift} follows given only that the $J^{2}$ terms of~\eqref{eq:thermo_full}-\eqref{eq:thermo_eps} are positive.

The physical picture is then transparent.
Switched on at fixed temperature, the deformation \emph{adds} entropy, energy and pressure: the operator acquires an expectation value $\langle\mathcal{O}_{\phi}\rangle=\chi J\neq0$ and the horizon grows.
If instead the plasma is prepared with a prescribed entropy--or equivalently a prescribed energy--that extra capacity must be paid for by a lower temperature, so the deformed plasma is \emph{cooler} than the conformal one carrying the same entropy.
In the energy language, part of the budget is stored in the condensate rather than in thermal excitations, leaving a smaller horizon and hence a smaller $T$.

Two features deserve emphasis.  First, the cooling is a UV effect: $\delta T/T\propto J^{2}T^{\sqrt{33}-3}$ with $\sqrt{33}-3=2(\De_{+}-3)\approx2.74>0$, so it grows with temperature
and switches off as $T\to0$.
This is the thermodynamic counterpart of $\zeta/s\to0$ and $\varepsilon-2p\to0$ in the infrared: the deformation is irrelevant, the IR fixed point is restored, and no trace of the hair
survives in the low-temperature thermodynamics.
At fixed source the trace itself follows the pure power law $J\langle\mathcal{O}_{\phi}\rangle\propto T^{2\De_{+}-3}=T^{\sqrt{33}}$.
Second, the cooling and the softening of the equation of state are two faces of the same response.
Both are governed by the single combination
$\chi_{0}J^{2}\rh^{\sqrt{33}-3}$, and
\begin{equation}\label{eq:cs2_T_link}
  c_{s}^{2}-\tfrac12 = (\De_{+}-3)\,\frac{\delta T}{T}\bigg|_{s},
\end{equation}
so the conformal dimension of the deforming operator alone fixes the ratio between them.

No $\mathcal{O}(\eps)$ shift arises in any ensemble.
Because $\phi_{0}$ is a critical point of $V$, the undeformed theory has $\langle\mathcal{O}_{\phi}\rangle=0$ and the free energy~\eqref{eq:Ffull} contains no term linear in $J$; every
thermodynamic response therefore begins at $J^{2}=\mathcal{O}(\eps^{2})$.

\subsubsection*{On the apparent zero of $c_{s}^{2}$}

Taken literally, the truncated expression~\eqref{eq:cs2_exact} reaches $c_{s}^{2}=0$ at $\eps\approx0.46$ and turns negative beyond.
Because a negative $c_{s}^{2}$ would be physically dramatic, it is worth stating precisely what our calculation does and does not establish.

A genuine $c_{s}^{2}<0$ would signal a local thermodynamic instability of the homogeneous phase.
With $c_{s}^{2}=(dp/dT)_{J}/(d\varepsilon/dT)_{J}$ and $s>0$, it requires a negative specific heat $(\partial\varepsilon/\partial T)_{J}<0$; equivalently the sound speed in
$w=\pm c_{s}k-\tfrac{i}{2}\Gamma_{s}k^{2}$ becomes imaginary, so the long-wavelength sound mode grows exponentially rather than propagating.
Such a spinodal region would indicate that the uniform brane is unstable against phase separation into inhomogeneous domains.

Our expansion cannot decide whether this happens.
At $\eps\approx0.46$ the $\mathcal{O}(\eps^{2})$ correction in~\eqref{eq:cs2_exact} equals $100\%$ of the conformal value $\tfrac12$, so the neglected $\mathcal{O}(\eps^{4})$ term is of the same size and the series is no longer controlled.
The ambiguity is explicit: organizing the \emph{same} $\mathcal{O}(J^{2})$ information without re-expanding the ratio, i.e.\ evaluating $(dp/dT)_{J}/(d\varepsilon/dT)_{J}$ directly
from~\eqref{eq:thermo_full}-\eqref{eq:thermo_eps}, gives a $c_{s}^{2}$ that decreases monotonically from $\tfrac12$ to $1/(\sqrt{33}-1)=(1+\sqrt{33})/32\approx0.211$ and never vanishes, since
$(dp/dT)_{J}$ and $(d\varepsilon/dT)_{J}$ are then manifestly positive sums.
Two equally legitimate treatments of the same data therefore disagree about the existence of the zero, which settles the matter: the sign of $c_{s}^{2}$ at $\eps\gtrsim0.46$ is not determined at this order.

We therefore quote $\eps\lesssim0.46$ as the range over which $c_{s}^{2}$ is quantitatively reliable--consistent with, and marginally sharper than, the thermodynamic bound $\eps\lesssim0.5$ of
Sec.~\ref{sec:disc}--and we make no claim of a spinodal instability.
Settling that question requires the finite-$\eps$ numerical solution discussed in Sec.~\ref{sec:disc}, which would also determine whether the hairy phase is thermodynamically preferred.

\section{Discussion}
\label{sec:disc}

\subsection{Summary of results}

We have constructed perturbative scalar-hairy black-brane solutions in the $G_{2}$-invariant ISO(7) supergravity to $\mathcal{O}(\eps^{2})$ in closed analytic form.
The principal results are collected in Table~\ref{tab:summary}.

\begin{table}[!htbp]
\centering
\renewcommand{\arraystretch}{1.45}
\small
\begin{tabular}{lll}
\toprule
Quantity & Formula & Ref.\ \\
\midrule
Horizon radius & $\rh=(2\om l^{2})^{1/3}$ & \eqref{eq:rh_def}\\
Hawking temperature & $T^{(0)}=3\rh/(4\pi l^{2})$
  & \eqref{eq:T0}\\
Entropy density & $s^{(0)}=\rh^{2}/(4G_{4})$
  & \eqref{eq:T0}\\
$\nu$ (scalar PT) & $\nu=-\tfrac{1}{2}+\tfrac{\sqrt{33}}{6}$,
  $\nu(\nu+1)=2/3$ & \eqref{eq:nu_explicit}\\
$\phi^{(1)}(r)$ & $P_{\nu}(2r^{3}/\rh^{3}-1)$
  & \eqref{eq:phi1sol}\\
$A^{(1)}(r)$ & $-\tfrac{\svn}{2}\,P_{\nu}(2r^{3}/\rh^{3}-1)$ (exact)
  & \eqref{eq:A1_exact}\\
$\mathcal{S}_{\phi}^{(2)}(r)$ & eq.~\eqref{eq:S2phi_r}
  & Sec.~\ref{sec:phi2}\\
$\mathcal{S}_{E}^{(2)}(r)$ & eq.~\eqref{eq:S2A_r}
  & Sec.~\ref{sec:A2}\\
Holographic identity & $\nu+1=\Delta_{+}/3$
  & \eqref{eq:identity}\\
$d_{Q}=e_{Q}=0$ (exact) & horizon regularity
  & Secs.~\ref{sec:phi2}--\ref{sec:A2}\\
$\langle\mathcal{O}\rangle/J$ unchanged & to $\mathcal{O}(\eps^{2})$
  & \eqref{eq:JO_num}\\
Trace / EoS & $\varepsilon-2p
  =(\De_{+}-3)\,J\langle\mathcal{O}_{\phi}\rangle/l^2$
  & \eqref{eq:trace_exact}\\
Speed of sound & $c_{s}^{2}=\tfrac12
  +\tfrac{\sqrt{33}(3-\sqrt{33})}{24}\,\chi_{0}J^{2}\rh^{\sqrt{33}-3}$
  & \eqref{eq:cs2_exact}\\
Temperature shift & $\delta T/T|_{s}
  =-\tfrac{\sqrt{33}}{12}\,\chi_{0}J^{2}\rh^{\sqrt{33}-3}$
  & \eqref{eq:Tshift}\\
$\eta/s$ (universal) & $1/(4\pi)$ & \eqref{eq:etas_bound}\\
Bulk viscosity & $\zeta/\eta=\tfrac{3(7-\sqrt{33})}{8}\eps^{2}$
  & \eqref{eq:zeta}\\
Shear diffusion & $D_{\perp}=1/(4\pi T)$ & \eqref{eq:Dperp}\\
Sound attenuation & $\Gamma_{s}=(1+\zeta/\eta)/(4\pi T)$ & \eqref{eq:Gammas}\\
\bottomrule
\end{tabular}
\caption{Summary of main results.
  Here $\hatP(r)=P_{\nu}(2r^{3}/\rh^{3}-1)$,
  and $P_{1}$, $Q_{1}$ are given by eqs.~\eqref{eq:P1r}--\eqref{eq:Q1r}.}
\label{tab:summary}
\end{table}

\subsection{Physical implications for the dual SCFT$_{3}$}

\paragraph{Bulk viscosity.}
The Eling-Oz formula~\eqref{eq:zeta_ward} gives a nonzero bulk viscosity that encodes the breaking of conformal invariance by
the scalar hair:
\begin{equation*}
  \frac{\zeta}{\eta} =\frac{3(7-\sqrt{33})}{8}\,\eps^{2}\approx 0.471\,\eps^{2}, \qquad
  \frac{\zeta}{s} = \frac{3(7-\sqrt{33})}{32\pi}\,\eps^{2} \propto T^{\sqrt{33}-3}\approx T^{2.74}.
\end{equation*}
The temperature dependence $\zeta/s\propto T^{\sqrt{33}-3}$ vanishes as $T\to 0$, confirming that conformal invariance is recovered in the infrared, and grows in the ultraviolet,
consistent with the deformation being irrelevant ($\Delta_{+}=(3+\sqrt{33})/2>3$).
The coefficient $3(7-\sqrt{33})/8\approx 0.471$ is a pure number fixed entirely by the ISO(7) supergravity data (specifically, by the scalar mass $m_{\phi}^{2}l^{2}=6$ at the $G_{2}$ critical point), and constitutes a prediction for the ratio $\zeta/\eta$ of the dual SCFT$_{3}$ at strong coupling.

\paragraph{Equation of state.}
The $\mathcal{O}(\eps^{2})$ corrections deform the conformal equation of state $\varepsilon=2p$ to
\begin{equation*}
  \varepsilon-2p = (\Delta_{+}-3)\,J\langle\mathcal O_{\phi}\rangle/l^2 \propto J^{2}T^{\sqrt{33}},
\end{equation*}
in agreement with the conformal Ward identity
$\varepsilon-2p = -\langle T^{\mu}_{\ \mu}\rangle = (\Delta_{+}-3)\,J\langle\mathcal O_{\phi}\rangle/l^2$.
The positive exponent $\sqrt{33}\approx 5.74$ implies that the trace anomaly grows rapidly with temperature, reflecting the increasing importance of the irrelevant coupling at high energies.

\paragraph{Holographic identity.}
The exact identity $\nu+1=\Delta_{+}/3$ provides a nontrivial consistency check linking the bulk P\"oschl-Teller parameter to the boundary conformal dimension of the dual operator.
This identity follows algebraically from $3(\nu+1)=3\nu+3=(\Delta_{+}-3)+3=\Delta_{+}$, but its origin is the nonlinear Fefferman-Graham map $z\sim z_{\mathrm{FG}}^{3}$ relating the domain-wall coordinate to the standard holographic coordinate.

\subsection{Effective conformal dimension and the role of $r_{h}$}
\label{sec:eff_dim}

The scalar potential at the $G_{2}$ critical point gives $V''(\phi_{0})l^{2}=6$, corresponding to an irrelevant operator of dimension $\Delta_{+}\approx 4.37$.
However, the second derivative of the potential is not sign-definite along the scalar flow.
Specifically,
\begin{equation}\label{eq:Vpp_flow}
  V''(\phi) = -\frac{5g_{c}^{2}}{2}\,e^{-\phi/\svn} + \frac{7\mR^{2}}{2}\,e^{-\svn\phi},
\end{equation}
which changes sign at $\phi_{*}=\phi_{0}+\frac{\svn\ln7}{6}$.
Since the first exponential decays more slowly than the second ($1/\svn<\svn$), for $\phi>\phi_{*}$ the effective mass squared becomes negative, $V''(\phi)<0$, and vanishes from below as
$\phi\to+\infty$.

On the fixed AdS$_4$-Schwarzschild background with AdS radius~$l$, the effective dimensionless mass squared is $m_{\rm eff}^{2}=V''(\phi)$, and the corresponding effective conformal dimension is $\Delta_{\rm eff}^{+} = \tfrac{3}{2}+\sqrt{\tfrac{9}{4}+V''(\phi)l^{2}}$.
Along the scalar profile $\phi^{(1)}(r)=P_{\nu}(u(r))$, which grows as $r\to\infty$, one finds:
\begin{center}
\renewcommand{\arraystretch}{1.4}
\begin{tabular}{cccc}
\toprule
Region & $V''(\phi)l^{2}$ & $\Delta_{\rm eff}^{+}$ & Character \\
\midrule
$\phi=\phi_{0}$ (horizon, IR) & $+6$ & $4.37$ & Irrelevant \\
$\phi=\phi_{*}$ (crossover) & $0$ & $3$ & Marginal \\
$\phi\to+\infty$ (boundary, UV) & $0^{-}$ & $3^{-}$ &
  Marginally relevant \\
\bottomrule
\end{tabular}
\end{center}
This transition reflects the structure of the ISO(7) scalar potential: at large field values the Romans mass term $\propto e^{-\svn\phi}$ is exponentially suppressed, leaving the single exponential $V(\phi)\approx-(35g_{c}^{2}/2)e^{-\phi/\svn}$, for which $V''(\phi)l^{2}\to0^{-}$.
Physically, the scalar operator is irrelevant at the $G_{2}$ fixed point but becomes marginally relevant in the UV, which is precisely what allows the scalar profile to grow toward the boundary.
The same UV pattern--an operator whose effective dimension approaches~$3$ from below with logarithmic running--is realized exactly, and nonperturbatively, in the companion analytic model
of~\cite{Yun:2026exact}.

\paragraph{Comparison with relevant-deformation models.}
In the more commonly studied \emph{relevant} deformation scenario~\cite{Gubser:2008ny,Gursoy:2007bu,Gursoy:2007za}, the scalar field interpolates between two critical points $\phi_{\rm UV}$ and $\phi_{\rm IR}$ with $V'(\phi)=0$ at both ends.
The scalar is small at the UV boundary and flows to a second fixed point in the deep interior, producing exact IR conformality: the near-horizon geometry is precisely AdS$_4$ with a (possibly different) radius.

In the present $G_{2}$ model, the perturbative flow considered here is built around a single critical point, $\phi_{0}$: unlike the relevant-deformation scenario above, there is no second, nearby fixed point that this flow connects to.
(The ISO(7) landscape does contain a second $G_{2}$-invariant critical point, the non-supersymmetric vacuum of Sec.~\ref{sec:nsusy}, but it is parametrically far from $\phi_{0}$ and plays no role in the perturbative construction here.)
The horizon scalar $\phi(r_{h})=\phi_{0}+\eps$ is close to $\phi_{0}$ but not exactly at it, and there is no second fixed point for the flow to reach.
As a consequence, the near-horizon geometry is only \emph{approximately} conformally invariant--the nonconformality is proportional to $\eps^{2}\propto J^{2}r_{h}^{\sqrt{33}-3}$, which vanishes at $T\to0$ but is nonzero at any finite temperature.

\paragraph{Dual role of $r_{h}$.}
This structure may appear counterintuitive: if the deformation is irrelevant, its effects should grow in the UV, yet the nonconformality (as measured by $\zeta$ or $\varepsilon-2p$) is encoded at the horizon, i.e., the geometric IR.
The resolution is that $r_{h}$ simultaneously plays two distinct roles.
First, $r=r_{h}$ is the geometric infrared endpoint of any given black-brane geometry.
Second, $r_{h}\propto T$ parameterizes the temperature, so larger $r_{h}$ means higher temperatures, which probe the UV regime of the dual field theory.
The horizon scalar charge $\eps\propto J\,r_{h}^{\Delta_{+}-3}$ grows with $r_{h}$ precisely because the irrelevant coupling becomes more important at higher energies.
Thus, the nonconformality observed at the horizon is the imprint of the UV physics of the irrelevant deformation on the horizon data.

\paragraph{Necessity of perturbation theory.}
Since the scalar grows toward the boundary as $\phi^{(1)}\sim r^{\Delta_{+}-3}$, the full nonlinear solution does not admit standard asymptotically AdS$_{4}$ boundary conditions for finite deformation.
The perturbative expansion in $\eps$ is therefore not merely a technical convenience: it is the physically appropriate framework for incorporating an irrelevant deformation, analogous to treating a non-renormalizable coupling in effective field theory order by order.
At each order in $\eps$, the perturbation fields are controlled expansions around the AdS$_{4}$-Schwarzschild background, the boundary conditions are well defined, and physical quantities (free energy, entropy, transport coefficients) receive finite, calculable corrections.

\subsection{Stability of the $\chi=0$ truncation}
\label{sec:chi_stability}

The $\mathbb{Z}_{2}$ symmetry $\chi\to-\chi$ of the full two-scalar potential makes $\chi=0$ an exact consistent truncation.
At the $G_{2}$ critical point, $m_{\chi}^{2}=m_{\phi}^{2}=6/l^{2}$ (required by the $\mathcal{N}=1$ SUSY Ward identity), which lies well above the BF bound $m_{\mathrm{BF}}^{2}l^{2}=-9/4$.
The truncation is therefore perturbatively stable.

Beyond the perturbative regime, a tachyonic instability in $\chi$ can develop for $\phi$ significantly displaced from $\phi_{0}$.
A numerical scan of the $\chi$-direction of the ISO(7) scalar potential places this threshold at $\phi_{\rm tach}\approx\phi_{0}+0.77$ (an estimate, quoted here only to set the scale; it lies below $\phi_{*}$, the crossover point of Sec.~\ref{sec:eff_dim}).
For small $\eps$, $\phi=\phi_{0}+\mathcal{O}(\eps)$ and this instability is automatically avoided.

The perturbative stability of the full $G_{2}$ SUSY vacuum (with all 70 scalars) was established by Guarino, Malek, and Samtleben~\cite{Guarino:2020ltm} using $G_{2}$ representation theory.

\subsection{The non-supersymmetric $G_{2}$ vacuum}
\label{sec:nsusy}

The $G_{2}$-invariant sector also contains a non-supersymmetric AdS$_{4}$ critical point at $\phi_{0}^{\rm nSUSY}\neq\phi_{0}$~\cite{Borghese:2012qm}.
Its perturbative stability was proved analytically in~\cite{Guarino:2020ltm}: all bosonic Kaluza-Klein scalar masses at this vacuum satisfy the BF bound, providing one of the very few known candidate counterexamples to the non-supersymmetric AdS swampland conjecture of Ooguri and Vafa~\cite{Ooguri:2016pdq}.
Additional stability evidence includes the absence of brane-jet instabilities~\cite{Guarino:2020bkb,Bena:2020} and positive-energy arguments~\cite{Dibitetto:2021esi}, although nonperturbative bubble nucleation remains an active area~\cite{Bomans:2021ara}.
Constructing hairy black branes around this non-SUSY vacuum would provide a holographic probe of its stability and is a natural extension of the present work.

\subsection{Outlook}

Several natural extensions of this work suggest themselves.

\paragraph{(i) Numerical finite-$\eps$ solutions.}
The perturbative expansion is expected to break down at $\eps\lesssim\mathcal{O}(1)$.
Four independent physical criteria provide quantitative estimates of the convergence radius: the $\chi$-field tachyonic instability at $\phi_{\rm tach}\approx\phi_{0}+0.77$ gives $\eps\lesssim 0.77$; the $V''$ sign change at $\phi_{*}=\phi_{0}+\svn\ln7/6\approx\phi_{0}+0.86$ gives $\eps\lesssim 0.86$; thermodynamic self-consistency ($\de s/s^{(0)}<1$) gives $\eps\lesssim 0.5$; and requiring the $\mathcal{O}(\eps^{2})$ correction to $c_{s}^{2}$ to stay below the conformal value $\tfrac12$ (Sec.~\ref{sec:observables}) gives $\eps\lesssim 0.46$.
The most stringent bound is therefore $\eps\lesssim 0.46$.
Moreover, for any $\eps>0$, the growing scalar profile $P_{\nu}(u(r))\sim r^{\De_{+}-3}$ reaches $\phi_{\rm tach}$ at a finite radius $r_{*}/\rh\approx 3.8$ ($\eps=0.1$) to $r_{*}/\rh\approx 1.3$ ($\eps=0.46$), beyond which the $\chi=0$ truncation becomes unreliable.
A full numerical construction~\cite{Hertog:2004ns} for finite $\eps$ would determine whether the hairy phase is thermodynamically preferred over the bald AdS$_{4}$-Schwarzschild phase, and would
clarify the interplay between the effective-dimension crossover of Sec.~\ref{sec:eff_dim} and the $\chi$-field instability of Sec.~\ref{sec:chi_stability}.

\paragraph{(ii) Quasi-normal mode spectrum.}
The P\"{o}schl-Teller background provides an analytically tractable arena for computing the QNM spectrum of the scalar operator~\cite{Policastro:2002se,Birmingham:2001pj}.
The exact solvability of the PT equation suggests that the QNM frequencies can likely be expressed in closed form, analogously to the BTZ result of~\cite{Birmingham:2001pj}.

\paragraph{(iii) Charged hairy branes.}
Including the dyonic electromagnetic charges of the ISO(7) theory would extend the solutions to finite chemical potential $\mu$, potentially exhibiting holographic superconductivity~\cite{Hartnoll:2008vx} or non-Fermi liquid behavior~\cite{Liu:2009dm}.
Static BPS black holes of the dyonic ISO(7) theory have been constructed in~\cite{Guarino:2017pkh}, and their microstates counted holographically via supersymmetric localization~\cite{Benini:2015eyy,Azzurli:2017kxo}; extending the present neutral hairy branes into this charged sector would connect the two lines of development.

\paragraph{(iv) Higher-order perturbations.}
Extending to $\mathcal{O}(\eps^{3})$ and beyond would allow a systematic resummation of the $\eps$-series and comparison with numerics.

\paragraph{(v) Non-SUSY vacuum hairy branes.}
As noted in Sec.~\ref{sec:nsusy}, hairy branes around the non-SUSY $G_{2}$ vacuum would constitute a holographic probe of the Ooguri-Vafa conjecture~\cite{Ooguri:2016pdq}.

\appendix

\section{Properties of Legendre Functions}
\label{app:legendre}

We collect here the properties of the Legendre functions $P_{\nu}(u)$ and $Q_{\nu}(u)$ used in the main text.
The functions satisfy the Legendre differential equation
\begin{equation}\label{eq:Legendre_eq}
  \frac{d}{du}\!\left[(u^{2}-1)\frac{d\Phi}{du}\right] = \nu(\nu+1)\,\Phi, \qquad u\in(1,+\infty).
\end{equation}

\paragraph{Series representations near $u=1$.}
\begin{align}
  P_{\nu}(u) &= 1 + \frac{\nu(\nu+1)}{2}\,(u-1) + \mathcal{O}\!\left((u-1)^{2}\right), \label{eq:P_series_1}\\
  Q_{\nu}(u) &= -\frac{1}{2}\ln\frac{u-1}{2} + \psi(1)-\psi(\nu+1) + \mathcal{O}\!\left((u-1)\ln(u-1)\right), \label{eq:Q_series_1}
\end{align}
where $\psi=\Gamma'/\Gamma$ is the digamma function.
In $r$ coordinates, $u-1=2(r^{3}-\rh^{3})/\rh^{3}$, so $Q_{\nu}\sim-\tfrac{1}{2}\ln(\frac{r-\rh}{\rh})+\mathcal{O}(1)$ near $r\to\rh$: this is the logarithmic divergence that enforces
$c_{Q}=d_{Q}=e_{Q}=0$ by horizon regularity.

\paragraph{Asymptotic expansion as $u\to+\infty$.}
\begin{equation}\label{eq:Pasy_full}
  P_{\nu}(u) = \frac{\Gamma(\nu+\tfrac{1}{2})}{\sqrt{\pi}\,\Gamma(\nu+1)} \bigl[(2u)^{\nu} + \mathcal{O}(u^{\nu-2})\bigr] + \frac{\Gamma(-\nu-\tfrac{1}{2})}{\sqrt{\pi}\,\Gamma(-\nu)}  \bigl[(2u)^{-\nu-1} + \mathcal{O}(u^{-\nu-3})\bigr].
\end{equation}

\paragraph{Wronskian.}
\begin{equation}\label{eq:Wron_u}
  W_{u}[P_{\nu},Q_{\nu}] = P_{\nu}\frac{dQ_{\nu}}{du} - Q_{\nu}\frac{dP_{\nu}}{du} = \frac{-1}{u^{2}-1}.
\end{equation}
In $r$ coordinates:
\begin{equation}
  W_{r}^{(\nu)}[P_{\nu},Q_{\nu}] = \frac{du}{dr}\cdot W_{u}^{(\nu)} = \frac{6r^{2}}{\rh^{3}}\cdot\frac{-1}{u^{2}-1} = -\frac{3\rh^{3}}{2r(r^{3}-\rh^{3})}.
\end{equation}
Note that this formula holds for \emph{any} $\nu$ since $W_{u}=-1/(u^{2}-1)$ is $\nu$-independent.

\paragraph{Special values.}
For our parameter $\nu=-\tfrac{1}{2}+\tfrac{\sqrt{33}}{6}$:
\begin{equation}
  P_{\nu}(1)=1, \quad \nu(\nu+1)=\frac{2}{3}, \quad 2\nu+1 = \frac{\sqrt{33}}{3}, \quad \psi(1)-\psi(\nu+1) \approx -0.5731.
\end{equation}

\section{Holographic renormalization of the free energy}
\label{app:holoren}

This appendix derives the counterterms underlying the free-energy density~\eqref{eq:Ffull} and the one-point function~\eqref{eq:J_exact} by standard holographic
renormalization~\cite{deHaro:2000vlm,Bianchi:2001kw,Papadimitriou:2004ap}.
The dual scalar operator is \emph{irrelevant} ($\Delta_{+}=(3+\sqrt{33})/2>3$), so its source mode grows toward the boundary; the renormalization is organized order by order in the source, consistently with the $\eps$-expansion of the main text.
Two features of the black brane simplify the analysis decisively: the boundary geometry is \emph{flat} and the deformation is \emph{homogeneous}. Every curvature- and derivative-type boundary counterterm therefore vanishes identically, and the entire counterterm action is fixed by a single ``fake superpotential'' $W(\phi)$.

\subsection*{Conventions made explicit}

The counterterm bookkeeping below rests on three conventions that are left implicit in the main text; we fix them here once and for all.

\paragraph{(1) Euclidean continuation and signs.}
Thermodynamics is computed from the Euclidean on-shell action $S_{E}$ at inverse temperature $\beta=1/T$, with $\mathcal{F}=-T\ln Z=S_{E}^{\mathrm{ren}}/(\beta V_{2})$ and $V_{2}=\int d^{2}x$.  Continuing the Lorentzian action $S=\frac{1}{16\pi G_{4}}\int\sqrt{-g}\,(R-\tfrac12(\partial\phi)^{2}-V)$ gives $S_{E}=-\frac{1}{16\pi G_{4}}\int\sqrt{g}\,(R-\tfrac12(\partial\phi)^{2}-V)$; on shell the trace relation $R=\tfrac12(\partial\phi)^{2}+2V$ collapses the bulk integrand to $V$.  Together with the Gibbons--Hawking and counterterm boundary terms this fixes the signs
\begin{equation}\label{eq:Fsigns}
  16\pi G_{4}\,\mathcal{F} = -\int_{r_{h}}^{r_{c}}\!\sqrt{g}\,V\,dr \;-\;2\big[\sqrt{\gamma}\,K\big]_{r_{c}} \;-\;\big[\sqrt{\gamma}\,W(\phi)\big]_{r_{c}},
\end{equation}
with $W(\phi)$ the fake superpotential of the next subsection.
The overall normalization is pinned by the undeformed limit, in which \eqref{eq:Fsigns} reproduces the AdS$_{4}$-Schwarzschild value $\mathcal{F}_{0}=-r_{h}^{3}/(16\pi G_{4}l^{2})$ [the $J^{0}$ term of~\eqref{eq:Ffull}].

\paragraph{(2) Fixed flat boundary metric; homogeneous source.}
The free energy is a functional of the scalar source at a \emph{fixed} boundary geometry: we hold the conformal boundary metric flat, $\gamma_{(0)ij}=\delta_{ij}$, and vary only the homogeneous source $J$ (constant along the $x^{i}$).
All boundary data are then $x$-independent, so $\mathcal{R}[\gamma_{(0)}]=0$ and $\partial_{i}\varphi=0$; every curvature- and derivative-type counterterm vanishes and $S_{\mathrm{ct}}$ reduces to
$\int\sqrt{\gamma}\,W(\phi)$.  This also defines the one-point function through $\vev{\Oc_{\phi}}=\delta S_{\mathrm{ren}}/\delta J$ at fixed $\gamma_{(0)}$ [eq.~\eqref{eq:VEV_final}], and is the precise sense in which the $\mathcal{O}(J^{2})$ metric back-reaction can contribute only $J^{2}$-analytic contact terms.

\paragraph{(3) Radial slicing, horizon normalization, and the temperature.}
We work at fixed horizon radius $r_{h}$ and take for the Hawking temperature its background value.
The temperature is the surface gravity of the deformed brane,
\begin{equation*}
  T=\frac{\kappa}{2\pi},\qquad
  \kappa=\frac{|g_{tt}'|}{2\sqrt{-g_{tt}\,g_{rr}}}\bigg|_{r_{h}},
\end{equation*}
a property of the near-horizon geometry rather than of the Euclidean volume $\sqrt{g}=r^{2}e^{2\delta E}$.
The near-horizon data are fixed by the exact first-order identity $E^{(1)}=A^{(1)}+\tfrac{\svn}{2}\phi^{(1)}=0$ [eq.~\eqref{eq:A1_exact}], valid for all $r$, and the second-order normalization $\phi^{(2)}(r_{h})=A^{(2)}(r_{h})=0$ [eq.~\eqref{eq:second_order_norm}], which place the horizon at $r_{h}$ with the scalar regular there and give $\delta E(r_{h})=0$.
With this identification the potentials of eqs.~\eqref{eq:thermo_full}-\eqref{eq:thermo_eps} follow from $\mathcal F$ and form a first-law-consistent set, $d\varepsilon=T\,ds$ and $\varepsilon+p=Ts$; accordingly $r_{h}$ and $T$ are used interchangeably at the order worked here, and the irrelevant deformation is dialed by $\eps\propto J\,r_{h}^{\De_{+}-3}$ at fixed temperature.
An independent evaluation of $\varepsilon$ from the renormalized boundary stress tensor would furnish a direct check of this identification through the Gibbs-Duhem relation $\varepsilon+p=r_{h}\,\partial_{r_{h}}p$, equivalently $\varepsilon_{(2)}=(\sqrt{33}-1)\,p_{(2)}$ -- the same coefficient that fixes the trace $\varepsilon-2p=(\sqrt{33}-3)\,p_{(2)}J^{2}$ in eq.~\eqref{eq:trace_exact}.

\subsection*{Fake superpotential and counterterms}

Write the fluctuation $\varphi=\phi-\phi_{0}$ about the $G_{2}$ critical point, so that $V(\phi)=-\tfrac{6}{l^{2}}+\tfrac{3}{l^{2}}\varphi^{2} + \tfrac16 V'''(\phi_{0})\,\varphi^{3}+\cdots$, with
$m_{\phi}^{2}=V''(\phi_{0})=6/l^{2}$ and $V'''(\phi_{0})\,l^{2}=V_{3}=-48/\svn$ (eq.~\eqref{eq:V3_num}).
For a homogeneous flow $ds^{2}=du^{2}+e^{2A(u)}\eta_{ij}dx^{i}dx^{j}$ the Hamiltonian constraint $d(d-1)A'^{2}=\tfrac12\varphi'^{2}-V$ and the scalar equation are solved by the first-order system
$\varphi'=W'(\phi)$, $A'=-\tfrac{1}{2(d-1)}W$, provided $W$ satisfies
\begin{equation}\label{eq:superpot}
  V(\phi)=\tfrac12\bigl(W'(\phi)\bigr)^{2} - \frac{d}{4(d-1)}\,W(\phi)^{2} \;\xrightarrow{\;d=3\;}\; \tfrac12\,W'^{2}-\tfrac38\,W^{2}.
\end{equation}
The renormalizing counterterm is the boundary integral of this superpotential,
\begin{equation}\label{eq:Sct_W}
  S_{\mathrm{ct}} = \frac{1}{16\pi G_{4}}\int_{z=\epsilon}\!\!d^{3}x\,\sqrt{\gamma}\; W(\phi),
\end{equation}
whose cosmological piece reproduces the Balasubramanian-Kraus term: $W(\phi_{0})=-2(d-1)/l=-4/l$ gives $S_{\mathrm{ct}}\supset-\tfrac{1}{8\pi G_{4}}\int\sqrt{\gamma}\,(2/l)$.
On a curved or inhomogeneous boundary~\eqref{eq:Sct_W} would be supplemented by the $\tfrac{l}{2}\mathcal{R}[\gamma]$ term and by derivative counterterms $\sim(\partial\varphi)^{2},\, \mathcal{R}[\gamma]\varphi^{2}$; all of these vanish here.

Expanding $W=W_{0}+\tfrac12 W_{2}\varphi^{2}+\tfrac16 W_{3}\varphi^{3} + \cdots$ (note $W'(\phi_{0})=0$) and matching~\eqref{eq:superpot} order by order fixes the coefficients.
The $\varphi^{2}$ matching gives $W_{2}^{2}+\tfrac{3}{l}W_{2}-\tfrac{6}{l^{2}}=0$; the root reproducing the growing \emph{source} asymptotics ($\varphi\sim e^{W_{2}l A}\sim z^{\Delta_{-}}$) is $W_{2}=-\Delta_{-}/l=(\Delta_{+}-d)/l=(\sqrt{33}-3)/(2l)$, so the scalar mass counterterm $S_{\mathrm{ct}}\supset\frac{1}{16\pi G_{4}}\int\sqrt{\gamma}\, c_{2}\varphi^{2}$ carries the coefficient
\begin{equation}\label{eq:W2}
  \boxed{\;c_{2}\equiv\frac{W_{2}}{2} = \frac{\Delta_{+}-d}{2l}=\frac{\sqrt{33}-3}{4l}\approx\frac{0.686}{l}\;}\,.
\end{equation}
(The opposite root $W_{2}=-\Delta_{+}/l$ corresponds to the normalizable mode and does \emph{not} renormalize the source divergence; and $c_{2}\to0$ as $\Delta_{+}\to d$, the correct massless limit.)
The $\varphi^{3}$ matching, $\tfrac16 V'''(\phi_{0})=W_{3}\bigl(\tfrac12 W_{2}-\tfrac18 W_{0}\bigr)$, yields in closed form
\begin{equation}\label{eq:W3}
  W_{3} = \frac{l\,V'''(\phi_{0})}{3\,(1-\Delta_{-})} = -\frac{\sqrt{33}+1}{\svn\,l}\approx-\frac{2.55}{l},
\end{equation}
i.e.\ a cubic counterterm $\frac{1}{16\pi G_{4}}\int\sqrt{\gamma}\,\tfrac16 W_{3}\varphi^{3}$.
Higher coefficients $W_{4},W_{5},\dots$ follow recursively from~\eqref{eq:superpot}; they are the ``higher counterterms'' required by the irrelevant operator.

\subsection*{Which counterterms enter at $\mathcal{O}(J^{2})$}

Since $\varphi\sim\varphi_{(0)}\,z^{\Delta_{-}}\propto J$ near the boundary, the scalar counterterm $W_{n}\varphi^{n}$ in~\eqref{eq:Sct_W} contributes at order $J^{n}$.
In the \emph{scalar} sector the cubic~\eqref{eq:W3} and all higher $W_{n}$ therefore first act at $\mathcal{O}(J^{3})$, so the scalar power-law divergences at $\mathcal{O}(J^{2})$--the leading $z^{-\sqrt{33}}$ and its descendants $z^{-\sqrt{33}+2k}$--are removed by the mass counterterm $W_{2}$ alone, evaluated on the full Fefferman-Graham data $\varphi(z)=\varphi_{(0)}z^{\Delta_{-}}(1+\mathcal{O}(z^{2})) + \varphi_{(+)}z^{\Delta_{+}}(1+\mathcal{O}(z^{2}))$; this is guaranteed by~\eqref{eq:superpot}, i.e.\ by $S_{\mathrm{ct}}$ reproducing the divergent part of the on-shell action in the scalar channel.

The scalar simultaneously back-reacts on the metric, and this sector requires more care.  The $\mathcal{O}(\eps^{2})$ response $E^{(2)}\equiv A^{(2)}+\tfrac{\svn}{2}\phi^{(2)}$ [the warp combination $C+2H$ solved in Sec.~\ref{sec:A2}, eq.~\eqref{eq:E2_def}] is sourced by
$(\phi^{(1)})^{2}\sim r^{2(\Delta_{+}-3)}$; the source-driven response therefore scales as $r^{2(\Delta_{+}-3)}=r^{\sqrt{33}-3}$, which stays below the homogeneous metric mode $r^{3}$ (since $\sqrt{33}-3<3$).
In a \emph{direct} evaluation of $S_{\mathrm{reg}}+S_{\mathrm{GH}}$ the metric back-reaction nonetheless generates its own tower of divergences at \emph{irrational} exponents set by $\Delta_{+}$, which are \emph{not} of the scalar form $z^{-\sqrt{33}+2k}$ and are \emph{not} cancelled by the cosmological counterterm $W_{0}\sqrt{\gamma}$ evaluated on the \emph{background} induced metric.
These are supplied by $\sqrt{\gamma}\,W_{0}$ evaluated on the back-reacted induced metric $\gamma$ (not on its background value), whose finite remainder is $J^{2}$-analytic; because the induced boundary metric is flat ($R[\gamma_{(0)}]=0$) and the source homogeneous, each such remainder is a contact term $\propto J^{2}$ (with integer powers of $T$) and hence affects only the analytic part of $\mathcal{F}$ (see ``Scheme dependence'' below).
For this reason we do not obtain $\mathcal{F}$ by integrating the bare action; we instead read the renormalized one-point function from the normalizable scalar mode and fix the $J^{2}$ term of~\eqref{eq:Ffull} through the quadratic response, which is blind to these gravitational contact terms.

\subsection*{Renormalized free energy and one-point function}

With $S_{\mathrm{ren}}=\lim_{\epsilon\to0}(S_{\mathrm{reg}} + S_{\mathrm{GH}}+S_{\mathrm{ct}})$ finite, the renormalized one-point function is read off from the normalizable mode,
\begin{equation}\label{eq:VEV_final}
  \vev{\Oc_{\phi}} = \frac{1}{\sqrt{\gamma_{(0)}}}\frac{\delta S_{\mathrm{ren}}}{\delta J} = \frac{(2\Delta_{+}-d)\,\varphi_{(+)}}{16\pi G_{4}} = \frac{\sqrt{33}\,\varphi_{(+)}}{16\pi G_{4}},
\end{equation}
the coefficient $2\Delta_{+}-d=\sqrt{33}$ being exactly the factor that appears in~\eqref{eq:J_exact}. With the mode amplitudes of the main text
the physical susceptibility is $\chi=\vev{\Oc_{\phi}}/J=\chi_{0}\,\rh^{\sqrt{33}}/(16\pi G_{4})$, with the pure $\Gamma$-ratio $\chi_{0}=\sqrt{33}\,C_{-}/(C_{+}4^{\sqrt{33}/3})\approx2.335$
(eq.~\eqref{eq:chi_exact}); this fixes the $J^{2}$ term of~\eqref{eq:Ffull} via the quadratic response $\mathcal{F}=\mathcal{F}_{0}-\frac{1}{2l^2} J\vev{\Oc_{\phi}}+\mathcal{O}(J^{4})$.
Consistently with the counterterm analysis above, the metric back-reaction ($A^{(2)},B^{(2)}$) affects $S_{\mathrm{ren}}$ at this order only through local, $J^{2}$-analytic contact terms; the entire nonanalytic $\mathcal{O}(J^{2})$ response is carried by $\vev{\Oc_{\phi}}$.

\subsection*{Scheme dependence}

Being fixed by local counterterms, the finite pieces of $W(\phi)$ leave a residual scheme ambiguity: the renormalized free energy is determined only up to \emph{contact} terms analytic in $J^{2}$ with integer powers of $T$.
The physical, nonanalytic contribution $\propto J^{2}T^{\,2\Delta_{+}-3}=J^{2}T^{\sqrt{33}}$--and with it the trace identity~\eqref{eq:trace_exact} and the transport coefficients~\eqref{eq:zeta}--\eqref{eq:cs2_exact}--cannot be affected by any local counterterm and is therefore unambiguous.

The same protection covers the gravitational subtractions noted above: their exponents are irrational and so never reach the marginal value $r^{0}$, and since $2\Delta_{+}-d=\sqrt{33}\notin 2\mathbb{Z}$ there is no conformal anomaly and no surviving logarithm.
No choice of local subtraction---scalar or gravitational--can therefore contaminate the nonanalytic $J^{2}T^{\sqrt{33}}$ result, which is exactly $-\tfrac12\,J\vev{\Oc_{\phi}}$.

\section*{Acknowledgments}
I am grateful to Kyung Kiu Kim for helpful discussions.

\end{document}